\documentclass[%linenumbers,
trackchanges,twocolumn]{aastex701}
\usepackage{graphicx}
\usepackage{multirow}
\usepackage{hyperref}
\usepackage{xcolor}

\newcommand{\todo}[1]{{$\blacksquare$~\textbf{\color{red}#1}~$\blacksquare$}}

\DeclareRobustCommand{\Figref}[1]{Fig.~\ref{#1}}
\DeclareRobustCommand{\Tabref}[1]{Tab.~\ref{#1}}
\DeclareRobustCommand{\Secref}[1]{Sec.~\ref{#1}}

\graphicspath{{figures/}}

\begin{document}

\title{A grid of fast-rotating, chemically-homogeneous, supernova and/or long-GRB progenitors}

\author[0000-0002-6718-9472]{M.~Renzo}
\affiliation{Steward Observatory, Department of Astronomy, University of Arizona, 933 N. Cherry Ave., Tucson, AZ 85721, USA}
\affiliation{Center for Computational Astrophysics, Flatiron Institute, 162 5th Avenue, New York, NY 10010, USA}
\email{mrenzo@arizona.edu}

\author[0000-0003-3115-2456]{O.~Gottlieb}
\affiliation{Department of Physics and Kavli Institute for Astrophysics and Space Research, Massachusetts Institute of Technology, Cambridge, MA 02139, USA}
\affiliation{Center for Computational Astrophysics, Flatiron Institute, 162 5th Avenue, New York, NY 10010, USA}
\email{ogot@mit.edu}

\author[0000-0003-2776-082X]{H.~S.~Chan}
\affiliation{ILA, University of Colorado and National Institute of Standards and Technology, 440 UCB, Boulder, CO 80309-0440, USA}
\affiliation{Department of Astrophysical and Planetary Sciences, University of Colorado, 391 UCB, Boulder, CO 80309, USA}
\affiliation{Center for Computational Astrophysics, Flatiron Institute, 162 5th Avenue, New York, NY 10010, USA}
\email{hschanastrophy1997@gmail.com}

\author[0000-0003-1012-3031]{J.~A.~Goldberg}
\affiliation{Department of Physics and Astronomy, Michigan State University, East Lansing, MI 48824, USA}
\affiliation{Center for Computational Astrophysics, Flatiron Institute, 162 5th Avenue, New York, NY 10010, USA}
\email{goldstar@msu.edu}

\author[0000-0002-2215-1841]{A.~Grichener}
\affiliation{Steward Observatory, Department of Astronomy, University of Arizona, 933 N. Cherry Ave., Tucson, AZ 85721, USA}
\email{\todo{agrichener@arizona.edu}}

\author[0000-0002-8134-4854]{K.~Sen}
\affiliation{Steward Observatory, Department of Astronomy, University of Arizona, 933 N. Cherry Ave., Tucson, AZ 85721, USA}
\email{ksen@arizona.edu}

\author[0000-0002-8465-8090]{N.~Shah}
\affiliation{Steward Observatory, Department of Astronomy, University of Arizona, 933 N. Cherry Ave., Tucson, AZ 85721, USA}
\email{neevshah@arizona.edu}

\author[0000-0002-5794-4286]{E.~Farag}
\affiliation{Department of Astronomy, Yale University, New Haven, CT 06511, USA}
\email{ebraheem.farag@yale.edu}

\author[0000-0002-8171-8596]{Matteo Cantiello}
\affiliation{Center for Computational Astrophysics, Flatiron Institute, 162 5th Avenue, New York, NY 10010, USA}
\affiliation{Department of Astrophysical Sciences, Princeton University, Princeton, NJ 08544, USA}
\email{mcantiello@flatironinstitute.org}

\begin{abstract}
  The understanding of the mechanism(s) by which massive stars collapse and possibly explode is rapidly maturing. Uncertainties in the structure of the stellar core at the onset of collapse are often dominant in numerical simulations, and a limited number of progenitor grids are available. This is especially true for explosions where rotation and magnetic fields play a significant or primary role.
  We present a grid of 113 single-star models with initial masses $M_{\rm ZAMS}=30-90\,M_{\odot}$ and initial rigid rotation $\omega_{\rm ZAMS}=0.5-0.99\,\omega_{\rm crit}$ computed at $Z=0.001$ with the open-source stellar evolution code \textsc{MESA}. We adopt a 128-isotope nuclear reaction network capable of following the weak reactions deleptonizing the core during and after silicon core burning. By construction, these models experience rotationally-induced chemically-homogeneous evolution, and reach the onset of collapse ($v_{\rm infall}\lesssim -300\,\mathrm{km\ s^{-1}}$) with large and structured amounts of angular momentum, possibly sufficient to form accretion disks on a proto-compact object.
  Therefore, these progenitor structures provide a homogeneous set of models with updated input physics and improved algorithmic accuracy to understand stellar explosions of (some types of) stripped-envelope supernovae, possibly jetted and/or broad-lined, collapsars or magnetar-powered, and/or long $\gamma$-ray bursts.
\end{abstract}
\keywords{
  \uat{Gamma-ray bursts}{629} ---
  \uat{Supernovae}{1668} ---
  \uat{Stellar rotation}{1629} ---
  \uat{Stellar evolution}{1599} ---
  \uat{Massive stars}{732} ---
  \uat{High Energy astrophysics}{739}
}

\section{Introduction}
\label{sec:intro}

Since the establishment of a connection between massive-star
populations and supernovae (SNe, \citealt{baade:34}), understanding
how massive stars' end their life remains a capstone problem in
astrophysics. How does the iron core collapse and, at least in some
cases, how does the collapse successfully revert into an explosion? In
the past century, the increased cadence, wavelength coverage,
and depth of time-domain astronomy have transformed this question into
a family of related problems, spanning a diversity of astronomical
transients \citep[e.g.,][]{cano:17}. The interest in the origin of
transients and compact objects is even more timely in the era of
large time-domain surveys (e.g., Rubin/LSST, \citealt{ivezic:19}) and
routine detections of gravitational-wave signals from compact-object mergers
\citep[e.g.,][]{gw150914, mandel:22, abac:25, GWTC-5}.

Recent progress in SN theory has produced unprecedented agreement
among numerical simulations of neutrino-radiative transfer of core
collapse \citep[e.g.,][]{janka:12, melson:15, burrows:25, nakamura:25,
  mezzacappa:26, chen:26}. For non-rotating stellar progenitors,
neutrino-driven turbulent convection, sustained by continued accretion
onto the shock, can lead to successful stellar explosions
\citep[e.g.,][]{kuroda:18, ott:18, burrows:25}. While limits of this
mechanism, including the role of rotation and magnetic fields remain
debated \cite[e.g.,][]{soker:24, gottlieb:25}, a consensus on the fact
that uncertainties in the stellar progenitor structure are dominant is
emerging \citep[e.g.,][]{woosley:02, chieffi:13, sukhbold:16,
  farmer:16, renzo:17, mueller:17, limongi:18, davis:19, limongi:20, laplace:21,
  woosley:21, fields:22, bruenn:23, schneider:24, limongi:24, boccioli:24,
  renzo:24, limongi:25, burrows:25, maltsev:25, griffiths:26, myers:26}.

The structure of the stellar core of a progenitor model at the onset
of collapse determines the initial conditions for expensive
multi-dimensional, multi-physics (and often multi-million-CPU
hours) simulations \citep[e.g.,][]{janka:12, mueller:16rev,
  takiwaki:16, vartanyan:21, gottlieb:25}. Specifically, the profile
-- of density, entropy, velocity, angular momentum, magnetic fields,
etc.,~as a function of Lagrangian mass coordinate or radius --
determines the dynamics of the infalling gas and the ``over-burden''
caused by the accretion of outer layers through a putative
nascent outflow \citep[e.g.,][]{bruenn:13, bruenn:16}. Therefore, it
is the structure of the collapsing progenitor that decides whether
a stellar model explodes or not; separately, whether it leaves behind
a neutron star (NS) or a black hole (BH), either directly or through a
proto-NS phase; and ultimately the properties of the outflows,
the associated transient(s), and final remnants
\citep[e.g.,][]{kuroda:18, ott:18, vartanyan:19, burrows:24,
  nakamura:25, burrows:25}.

A current bottleneck in numerical simulations of stellar explosions
is the lack of models of stellar progenitors computed
\emph{i}) sufficiently late into the collapse and \emph{ii}) including
sufficient (nuclear and weak) physics to capture the known critical
ingredients for the explosion. This is especially true for fast
rotating progenitors of explosions where magnetic fields and angular
momentum may play a dominant role.

Specifically, condition \emph{i}) requires computing stellar evolution
models well into the sub-sonic phase of the collapse, at least until
the remaining time to core bounce becomes shorter than the
characteristic timescales of all relevant stellar processes. This
typically means evolving a stellar model beyond the point where the
iron-core infall velocity $v_{\rm infall} = \min(v)$ is lower than
$\lesssim -300\,\mathrm{km\ s^{-1}}$ \citep[e.g.,][]{farmer:21,
  farmer:23, gottlieb:24} if not $-1000\,\mathrm{km\ s^{-1}}$
\citep[e.g.,][]{woosley:02}. Beyond this point, matter typically
becomes partially optically thick to neutrinos, and the thermodynamic
quantities exceed the bounds of tabulated equations of state in
stellar evolution codes. However, already before the beginning
  of the collapse, several inherently multidimensional processes may
  play a critical role (e.g., convection, and especially rotation and
  magnetic fields \citealt{yadav:20, fields:22, chen:26}). For initial
  conditions to simulate these in multiple dimension, criterion
  \emph{i}) may need modifications to earlier times.

For condition \emph{ii}), stellar progenitor models used to study the
explosion physics must accurately capture the progenitor's density and
angular momentum structure. Any star evolving through silicon core
burning will experience copious amounts of electron captures
\cite[e.g.,][]{arnett:77,hix:93,hix:99,grichener:25, myers:26} which
progressively de-leptonize the core (i.e., decrease the amount of free
electrons). The notable exceptions to this are ``electron-capture'' SN
progenitors which collapse with an ONeMg-rich core (however, these too
require appropriate treatment of the weak reactions, including URCA
processes, \citealt{nomoto:84, takahashi:13, jones:14, zha:19}) and
progenitors above the pair-instability gap \citep[e.g.,][]{rakavy:67,
  renzo:24}, which collapse because of the photodisintegration
instability in an oxygen-rich core \citep[e.g.,][]{bond:84,
  gottlieb:25}. In both these cases, the stars reach their demise
bypassing the hydrostatic silicon core-burning phase.

In all other progenitors, electron captures occur at a significant
rate even during the hydrostatic silicon core burning phase, %and
before a model is sufficiently evolved according to \emph{i}). Even
adopting a ``quasi-statistical equilibrium'' approximation, the weak
reactions that determine $Y_{e}$ do not reach detailed balance, since
most occur when matter is still transparent to neutrinos, and the
equilibrium itself depends on $Y_{e}$ \citep[e.g.,][]{hix:96}.
Capturing the weak interactions in sufficient detail to characterize
the stellar \emph{structure} at the onset of collapse requires
considering $N\sim100$ isotopes \citep[e.g.,][]{farmer:16,
  grichener:25}. To decrease the computing time required to calculate
a grid of stellar progenitors, a commonly adopted shortcut is to use
small ($N\sim$20-isotope) nuclear reaction networks
\citep[e.g.,][]{pols:95, timmes:00, griffiths:25}. These bypass the
complex nuclear and weak physics of silicon core burning by lumping
all weak interactions in a custom compound reaction (see
\citealt{timmes:00, marchant:19, grichener:25}), effectively
sacrificing the accuracy of the core structure that decides the
explosion. Even if such a compound reaction is tuned on the central
value of the electron fraction $Y_{e, c}$ from models computed with
larger nuclear reaction networks
\citep[e.g.,][]{aguilera-dena:18, schneider:21, schneider:24}, small
networks cannot predict the \emph{profile} of the electron fraction
$Y_{e}(m)$ in the collapsing core where the explosions are determined
\citep[e.g.,][]{renzo:24}.

Using small nuclear reaction networks is a useful trade-off to study,
for example, electromagnetic observables of the progenitor or
explosions determined in the outer surface layers
\citep[e.g.,][]{morozova:15,paxton:18,goldberg:19,gilkis:25} or for
the application of semi-analytic explodability criteria based on
existing detailed simulations \citep[e.g.,][]{mueller:16,schneider:24,
  laplace:25,gilkis:26}. However, such networks are not sufficient for
investigations of explosion driving, pre-explosion and explosive
neutrino signals, nuclear yields, and the resulting compact-object
properties. This applies to both non-rotating \citep{farmer:16} and
rotating \citep{renzo:24} progenitors. The reason is that the
effective Chandrasekhar mass of the collapsing core scales with
$\propto Y_{e}^{2}$, or in other words, the contribution of
electron-degeneracy-pressure to the support of the core amplifies
small imprecision in the core structure to a level crucial for the
collapse and explosion dynamics \citep[e.g.,][]{janka:12,
  grichener:25}.

Several grids of stellar progenitors evolved sufficiently late
(condition \emph{i}) and with \emph{detailed} nuclear physics
(condition \emph{ii}) exist, mostly from closed-source codes (e.g.,
\textsc{KEPLER} \citealt{weaver:78, woosley:02, sukhbold:16,
  sukhbold:18} -- \textsc{FRANEC}\footnote{which does not include
  hydrodynamics and thus cannot reach condition \emph{i}), thus
  requiring an extra step to be used as progenitors in
  multi-dimensional simulations
  \citep[e.g.,][]{oconnor:10,limongi:20}.} \citealt{limongi:03,
  chieffi:13, limongi:18, limongi:24, limongi:25} -- \textsc{HOSHI}, \citealt{takahashi:13}). While these
stellar models have been invaluable in allowing the progress made so
far, challenges remain in the exploration of how the results depend on
the underlying \emph{physical} and \emph{numerical} choices in
modeling the evolution of stellar progenitors. Approximations at both
levels have been shown to affect the core structure and thus the
explosion outcome \citep[e.g.,][]{farmer:16, renzo:17, davis:19,
  laplace:21, farmer:21, farmer:23, maltsev:25, myers:26}. The few
public grids of progenitors fulfilling both conditions \emph{i}) and
\emph{ii}) computed with the open-source and community-driven code
\textsc{MESA} \citep[e.g.,][]{farmer:16, renzo:17, farag:20,
  laplace:21, farmer:21, farmer:23, myers:26} remain largely
unexplored in explosion simulations \citep[see
however][]{perna:18, vartanyan:21, gottlieb:24, chan:26}.

Despite the importance of fast-rotating progenitors for jetted
explosions, collapsars, magneto-hydrodynamic-explosions, and/or long
$\gamma$-ray burst (LGRB,
\citealt{woosley:93,macfadyen:94,gottlieb:25b}), even fewer grids of
such stellar models exist
\cite[e.g.,][]{heger:05,hirschi:04,hirschi:05,hirschi:05b,hirschi:05c}. To the best of our knowledge, these are
limited to the \textsc{KEPLER} models from \cite{woosley:06}, which
use the adaptive nuclear reaction network from \cite{heger:00}, and
the \textsc{MESA} models of \cite{aguilera-dena:18, aguilera-dena:20},
which adopt a small nuclear reaction network that cannot fulfill
condition \emph{ii}). Alternatively, non-rotating stellar models
endowed with ad-hoc angular momentum and magnetic field profiles that
are not consistent with the progenitor evolution are often used
\cite[e.g.,][]{siegel:22, gottlieb:25}.

The reason for the small number of stellar progenitor grids is the
substantial computational effort needed to evolve massive stars with large
nuclear-reaction networks\footnote{Neglecting sparsity, the size of the matrix to solve
  scales as $\sim{}N^{2}$, see Appendix B in \citealt{renzo:17} for
  the exact scaling in \textsc{MESA}.}, coupled with the stiffness of
the equations (see \citealt{grichener:25}), which results in large
fractions of progenitor stars encountering numerical issues before
reaching the desired infall velocity. These can manifest at the
surface or in the core (or both), and can often be an unclear mixture of
poorly-resolved physical phenomena and purely algorithmic problems.
Development of numerical strategies to clarify and address these issues are
ongoing \citep[e.g.,][]{paxton:13, jermyn:23,
  grichener:25,griffiths:25}.

In the meantime, we present here a publicly available grid of
fast-rotating models fulfilling both criteria \emph{i}) and \emph{ii})
for studies of the collapse and possible explosion of fast-rotating
stellar cores. We focus here on the overall grid
  of progenitor models and refer interested readers to \cite{renzo:24}
  and the Appendix in \cite{gottlieb:24} for more details of one
  specific example star.

In \Secref{sec:methods} we describe our physical
assumptions and numerical setup. \Secref{sec:results} presents the
grid of progenitors, and in \Secref{sec:scenario} we briefly outline
the astrophysical scenarios that they could approximate. We
contextualize our result in terms of the input and code physics in
\Secref{sec:discussion_inputs} before concluding in
\Secref{sec:conclusion}. Appendix~\ref{sec:mesa128} lists the 128
isotopes we track in our models, Appendix~\ref{sec:table} provides
some key quantities for the ``explodability'' and surface composition
at the onset of collapse. Appendix~\ref{sec:public_datafiles}
describes the input and output files of our simulations and our
processing scripts publicly available at
\href{https://doi.org/10.5281/zenodo.14286306}{doi.org/10.5281/zenodo.14286306}.

\section{Numerical setup}
\label{sec:methods}

The grid of models presented here is an extension of (and includes)
the progenitor star used in \cite{gottlieb:24} and constitutes the
input for the ``averaged'' progenitor constructed in \cite{chan:26}.

We use the \textsc{MESA} code (version \texttt{r24.03.1},
\citealt{paxton:11,paxton:13,paxton:15,paxton:16,paxton:19,jermyn:23}),
to compute low-metallicity ($Z=0.001$) and fast-rotating stellar
models. Our grid spans initial masses $M_{\rm ZAMS}=30-100\,M_{\odot}$
and initial (rigid) rotation with angular frequency
$\omega_{\rm ZAMS}=0.5-0.99\,\omega_{\rm crit}$, where
$\omega_\mathrm{crit}=\sqrt{(1-L/L_\mathrm{Edd})GM/R^3}$ is the
surface critical rotation rate accounting for radiative forces, $L$ is
the stellar luminosity, $L_\mathrm{Edd}$ the Eddington luminosity, $G$
is the gravitational constant, $M$ is the total mass and $R$ the
stellar radius. We focus on fast-rotating models to minimize numerical
issues in the envelopes \citep[e.g.][]{joss:73, paxton:13} and produce
sufficiently fast rotating core-collapse progenitors for explosions
studies where angular momentum and magnetic field play a role.

We use the \textsc{Skye} equation of state \citep{jermyn:skye}, and
include electron screening following \citet{Chugunov2007} and thermal
neutrino losses following \citet{Itoh1996}. We adopt radiative
opacities from OPAL \citep{Iglesias1993, Iglesias1996}
and electron conduction opacities from \citet{Cassisi2007}.

Nuclear reaction rates are a combination of NACRE \citep{Angulo1999},
JINA REACLIB \citep{Cyburt2010}, and
$^{12}\mathrm{C}(\alpha,\gamma)^{16}\mathrm{O}$ from \cite{kunz:02},
plus additional tabulated weak reaction rates \citet{Fuller1985,
  Oda1994, Langanke2000}. To capture with sufficient accuracy the
deleptonization of the core from the structural point of view
(condition \emph{ii}) in \Secref{sec:intro}), we adopt the
fully-coupled 128-isotope nuclear reaction network
\texttt{mesa\_128.net} \citep[see Appendix.~\ref{sec:mesa128}
and][]{farmer:16}. For central temperatures above
$\log_{10}(T)\ge 9.2$, we employ \textsc{MESA} in operator split mode,
that is we iteratively solve for the structure at fixed composition,
and vice versa (e.g., Sec.~10 in \citealt{jermyn:23}, see also
  \citealt{farmer:16, limongi:24, limongi:25}).

Throughout the evolution, we include the wind mass loss rate following
\cite{vink:00} unless the surface $^{4}\mathrm{He}$ mass
fraction $Y_{\rm surf}\le0.4$, in which case we adopt the rate from
\cite{nugis:00}. In both cases we assume the metallicity scaling from
\cite{vink:01} and the rotational enhancement from \cite{friend:86}
\citep[see also][]{langer:98}.

We use the Schwarzschild criterion and the \cite{henyey:65}
formulation of mixing-length theory for convective layers, with
$\alpha_{\rm MLT}=1.5$. \cite{anders:22} showed that for times
  long compared to the convective turnover timescale, convective
  penetration leads to the growth of the core to the
  Schwarzschild-defined boundary even starting from the
  \cite{ledoux:47}-boundary. For convective boundary mixing, we
include core overshooting as a step-function following \cite{brott:11}
($\alpha_{\rm ov}=0.335$) and treat rotational mixing in the diffusion
approximation following \cite{heger:00}, including angular momentum
transport by convection or by a ``classical'' Tayler-Spruit dynamo
\citep{spruit:02} in radiative regions.

All the models presented here result by construction in a
rotationally-driven chemically homogeneous evolution \cite[CHE,
e.g.,][]{maeder:00, yoon:06}. This limits (but does not eliminate) the
number of numerical issues occurring in the envelope. After the
formation of a carbon-oxygen core, we prevent the development of
spurious numerical velocities in the outer layers. We artificially set
to zero the velocity in any layer with sound-propagation time to the
surface longer than the current timestep, either from the outer edge
of this core, or the location where the specific entropy drops below
$s\le10.5\,k_{B}N_{A}$, whichever occurs at smaller radius.
Similar artificial damping exists in most calculations of
post-core-carbon burning massive stars
\citep[e.g.,][]{aguilera-dena:18}.

To produce progenitors sufficiently close to core-bounce (condition
\emph{i}) in \Secref{sec:intro}), we evolve each model until the
infall velocity in the iron core decreases below
$v_{\rm infall}\le-300\,\mathrm{km\ s^{-1}}$
\citep[e.g.,][]{farmer:23, gottlieb:24, myers:26}. This threshold is
motivated by the numerical experiments performed in
\cite{gottlieb:24}, where we showed that the physical time remaining
between this point and the canonical threshold of
$-1000\,\mathrm{km \ s^{-1}}$ is much shorter than the timescale
governing the local evolution of the rotation and magnetic
fields. % We re-implement the checks on the infall velocity in
% our \texttt{run\_star\_extras.f90} to avoid known bugs (already fixed
% in newer \textsc{MESA} versions).

Spatial and temporal resolution should always be tested in numerical
models of stellar evolution \citep[e.g.,][]{farmer:16}. We refer
readers to the appendix of \cite{gottlieb:24} for resolution tests
performed with the same setup and code on the initially
$40\,M_{\odot}$ and $\omega=0.6\,\omega_{\rm crit}$ model.

\section{Results}
\label{sec:results}

\begin{figure}[tp]
  \includegraphics[width=0.47\textwidth]{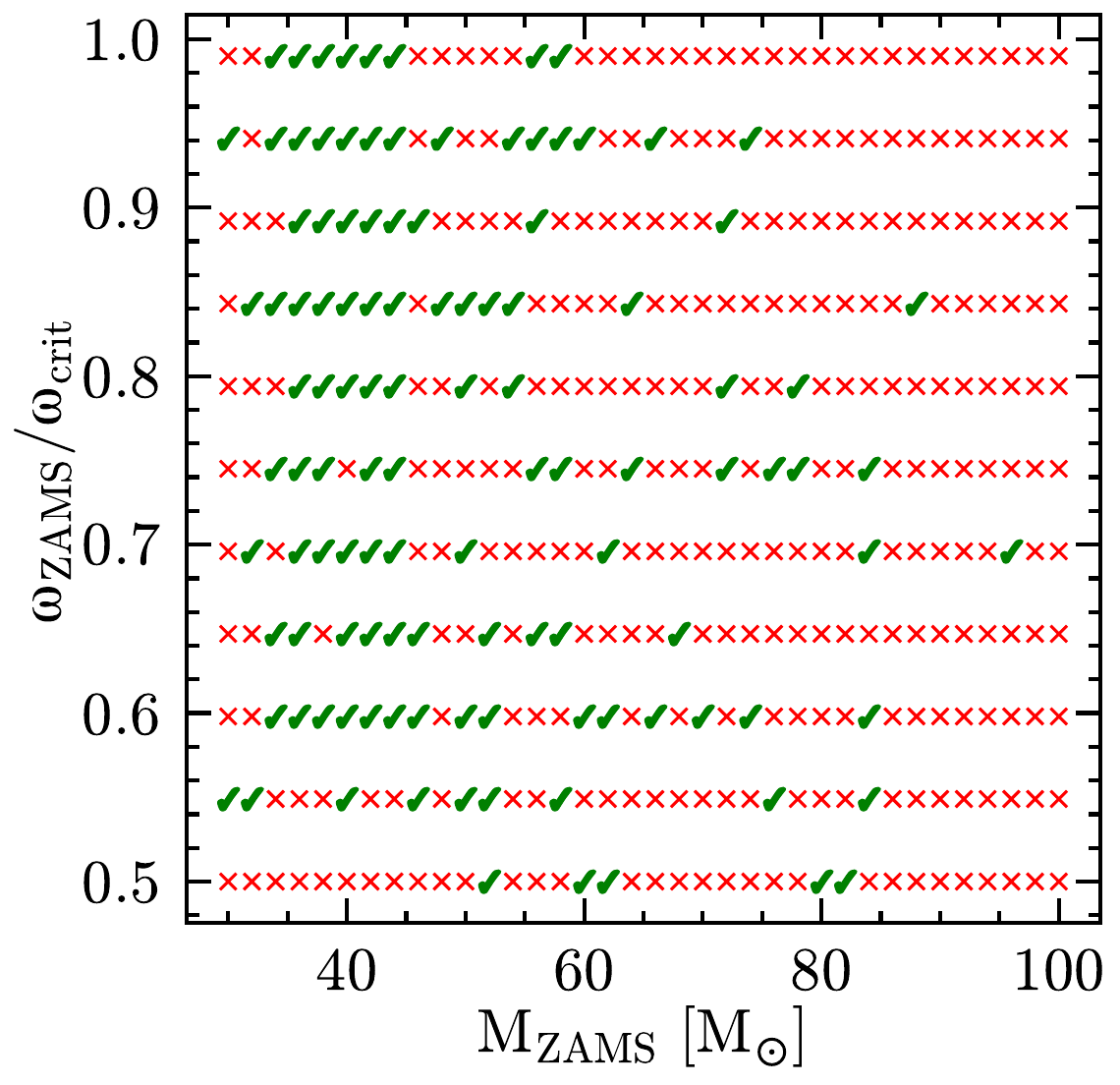}
  \caption{Overview of the grid success rate. Green checkmarks
    correspond to successful models. Red crosses correspond to models
    that either do not reach infall velocities of
    $-300\,\mathrm{km\ s^{-1}}$ or they exhibit large luminosity and
    effective temperature excursions that are treated unphysically in
    our setup and can impact the core-structure.}
  \label{fig:grid_success}
\end{figure}

Our grid consists initially of 11
values of $\omega_{\rm ZAMS}/\omega_{\rm crit}$ times 71
$M_{\rm ZAMS}$ values, for a total of 781 stellar models
(\Figref{fig:grid_success}). Of these, 113 successfully reach the
onset of core collapse and pass our visual inspection to remove models
experiencing large $L$ and effective temperature ($T_{\rm eff}$)
excursions. These may be manifestations of physical processes in the
envelope (e.g., \citealt{heger:97,fuller:17} for red supergiants and
e.g, \citealt{fuller:18} for envelope-less stars) and/or numerical
artifacts amplified by the small timesteps necessary in late
evolutionary phases (since the acceleration term in the momentum
equation is calculated as $\Delta v/\Delta t$, it can become extremely
large for small timesteps $\Delta t$, \citealt{paxton:11}).

Because of the fast initial rotation and assumed rotational mixing,
our models do not develop extended hydrogen envelopes, and our setup
is not designed to properly capture physical phenomena in the outer
layers this late in the evolution. Furthermore, large $L$ and
$T_{\rm eff}$ excursions feed back nonlinearly on the core in
structurally significant ways, motivating our choice to exclude models
exhibiting this numerical behavior from our grid. Overall, our grid
has a 14\% success rate, relatively high for models that are
\emph{not} manually mentored in any way.

As shown in \Figref{fig:grid_success}, successful models in our grid (green checkmarks) sample relatively well, the parameter space
$30\,M_{\odot}\lesssim M_{\rm ZAMS}\lesssim 50\,M_{\odot}$ across all
rotation rates explored
($0.5\lesssim \omega_{\rm ZAMS}/\omega_{\rm crit}\lesssim 0.99$), and
the number of successes decreases at larger initial masses and rotation
rates. However, there is no particular threshold beyond which models
systematically fail: this suggests most issues are related to
numerical errors in the solution of the stiff equations rather than
physical hypothesis.

\begin{figure}[bp]
  \includegraphics[width=0.5\textwidth]{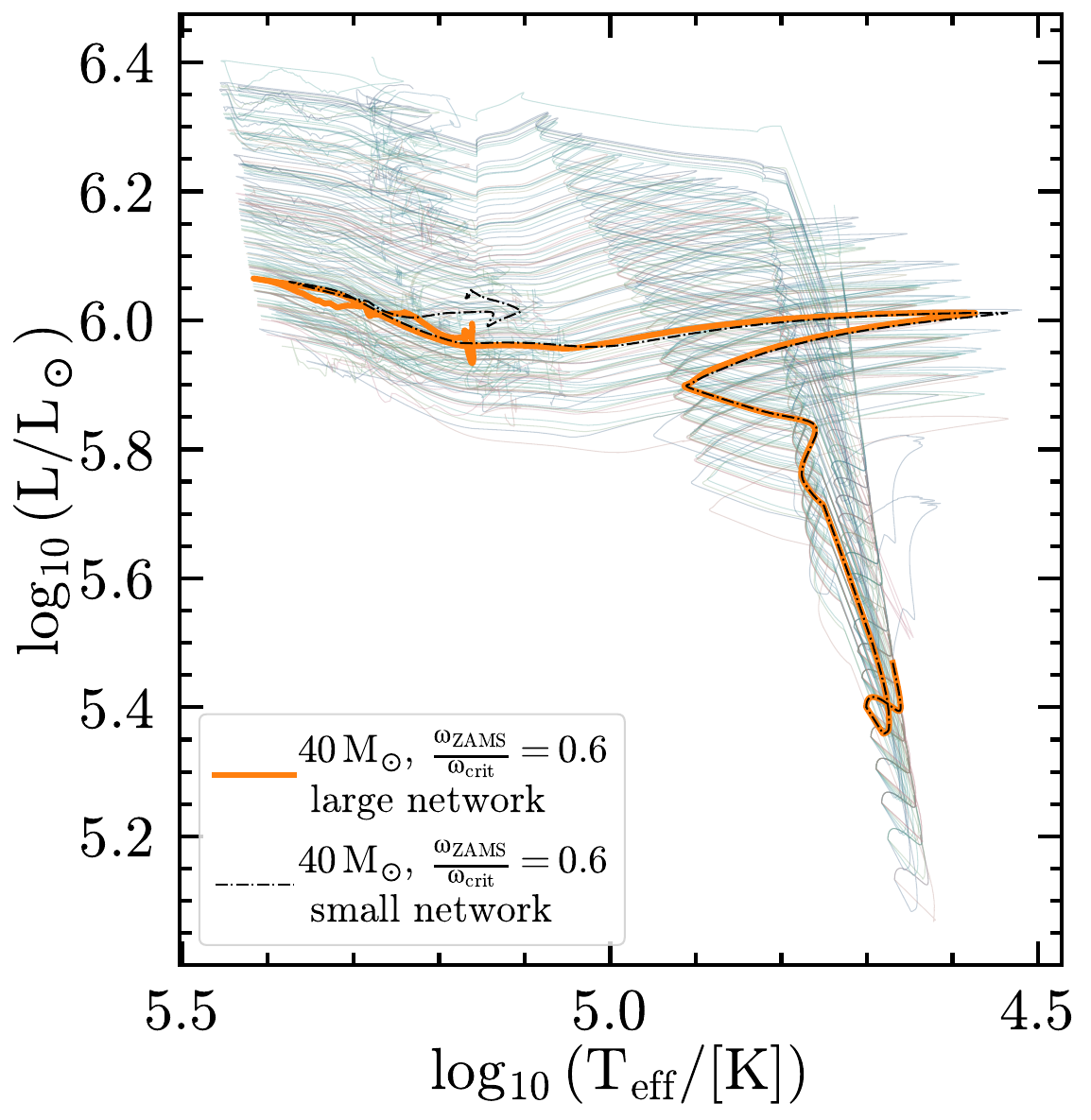}
  \caption{Herzsprung-Russel diagram of rotationally-induced
    chemically-homogeneous evolving single star models. The
    orange thick line marks a representative $40M_{\odot}$,
    $\omega_{\rm ZAMS}=0.6\,\omega_{\rm crit}$ used in
    \cite{gottlieb:24}, and the thin black dot-dashed line shows an
    otherwise identical model computed with only 22-isotopes from
    \cite{renzo:24}. Other colors indicate
      $\omega_{\rm ZAMS}/\omega_{\rm crit}$ as in
      Fig.~\ref{fig:xi_M}.}
  \label{fig:HRD}
\end{figure}

\begin{figure*}[htbp]
  \centering
  \includegraphics[width=\textwidth]{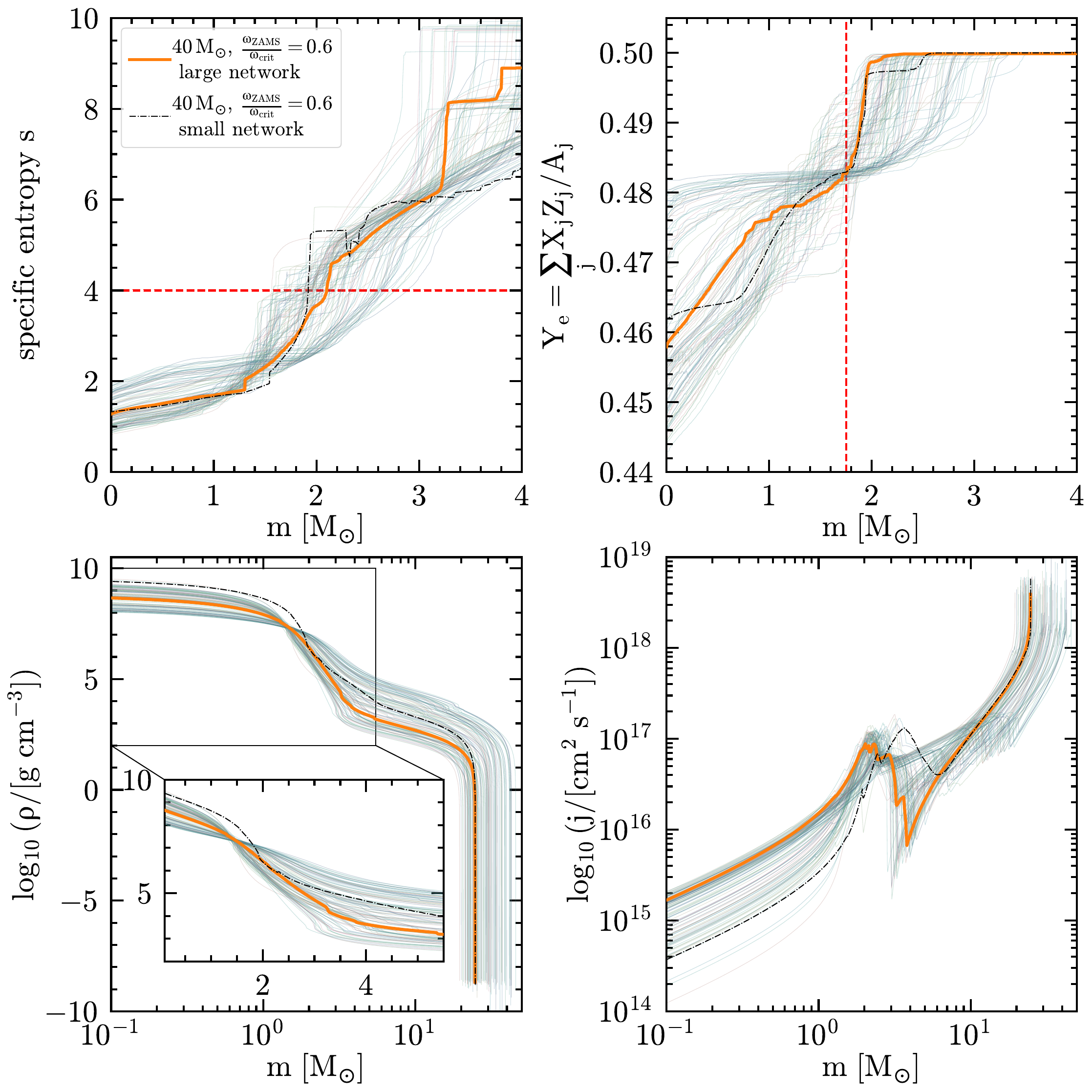}
  \caption{\emph{Top left}: Inner specific entropy profiles. The red
    dashed horizontal line marks the $s=4\,k_{B}N_{A}$ location often
    used in ``explodability'' criteria \cite[e.g.,][]{ertl:16,
      ertl:20}. \emph{Top right:} Inner electron fraction $Y_{e}(m)$
    profile as a function of mass. The red dashed vertical line marks
    $\mathcal{M}=1.75\,M_{\odot}$ used to calculate the compactness
    $\xi_{1.75}$. \emph{Bottom left:} Density profiles as a function
    of Lagrangian mass coordinate $m$ on a logarithmic scale. The
    inset zooms on the inner region and uses a linear scale.
    \emph{Bottom right:} Specific angular momentum profiles as a
    function of $m$. In each panel, the orange thick line marks a
    representative $40M_{\odot}$,
    $\omega_{\rm ZAMS}=0.6\,\omega_{\rm crit}$ used in
    \cite{gottlieb:24}, and the thin black dot-dashed line shows an
    otherwise identical model computed with only 22-isotopes from
    \cite{renzo:24}. Other colors indicate
      $\omega_{\rm ZAMS}/\omega_{\rm crit}$ as in
      Fig.~\ref{fig:xi_M}.}
  \label{fig:multipanel}
\end{figure*}

\begin{figure}[tbp]
  \includegraphics[width=0.5\textwidth]{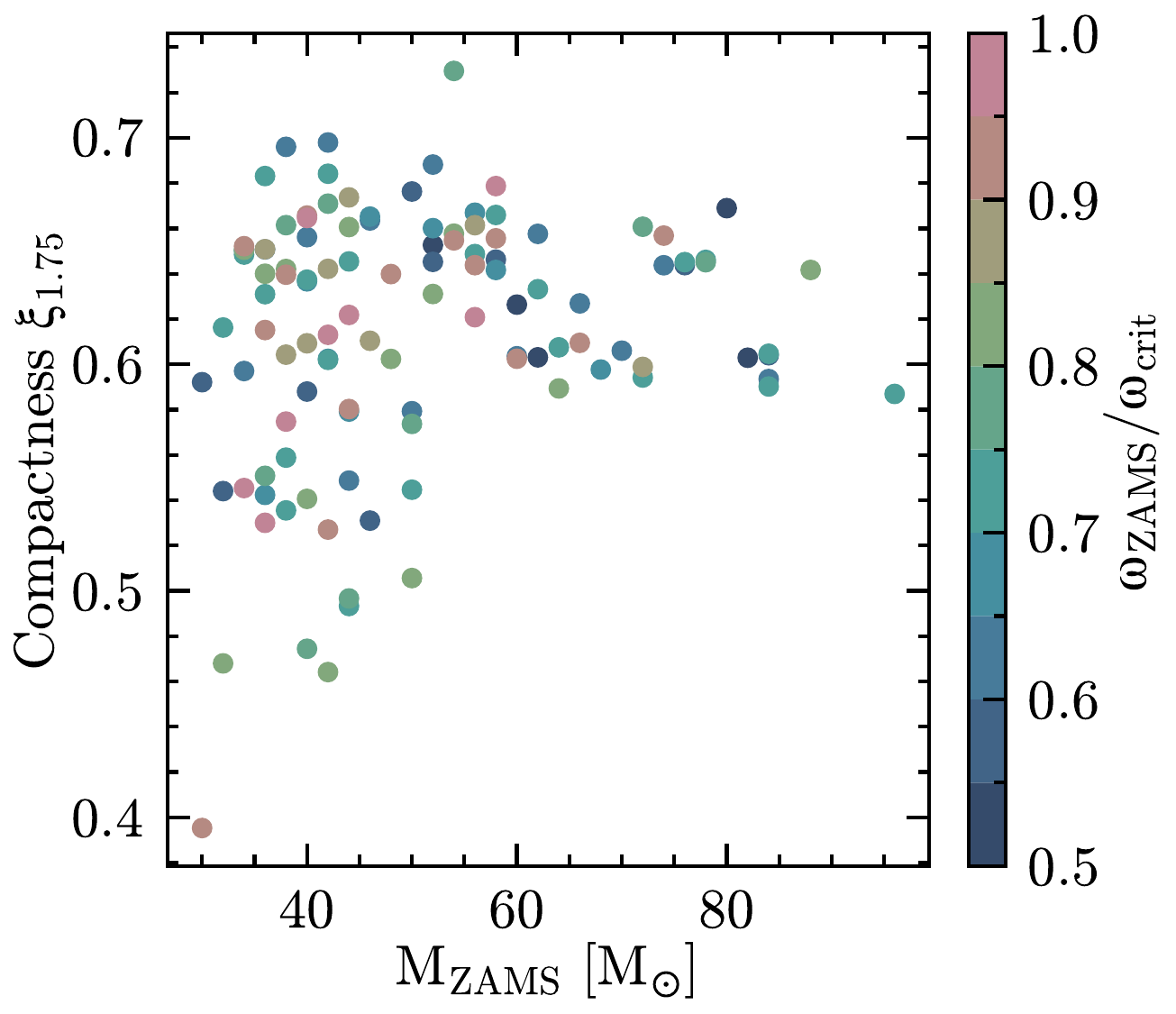}
  \caption{Compactness as a function of initial mass and rotation rate
    for single CHE models. Higher values of compactness
    $\xi_{\mathcal{M}}$ favor BH formation, but this does not imply
    without outflows and/or electromagnetic transients
    \citep[e.g.,][]{burrows:25, chan:26}. Colors
      indicate $\omega_{\rm ZAMS}/\omega_{\rm crit}$.}
  \label{fig:xi_M}
\end{figure}

From this point onward, we focus on the analysis of the successful
models. \Figref{fig:HRD} shows the evolution on the Herzsprung-Russel
diagram of our rotationally-induced CHE models with each thin line
corresponding to one star. Colors indicate the initial
$\omega_{\rm ZAMS}/\omega_{\rm crit}$, with darker colors
corresponding to values close to 0.5 and lighter colors for faster
initial rotation (see also color bar in \Figref{fig:xi_M}). The thick
orange lines correspond to the $40\,M_{\odot}$,
$\omega_{\rm ZAMS}=0.6\,\omega_{\rm crit}$ model used as progenitor in
\cite{gottlieb:24}, highlighted as a guideline. The thin black
dot-dashed line is the same model computed with a 22-isotope network
from \cite{renzo:24} available from
\href{https://doi.org/10.5281/zenodo.11375522}{doi.org/10.5281/zenodo.11375522}.

The stars evolve from the bottom right toward the top left. Because of
the large rotation imposed at birth, these models experience CHE
(\citealt{maeder:00, yoon:06, demink:09}). Rotational mixing
(dominated by meridional circulations) prevents the formation of a
chemical stratification and the emergence of a core--envelope boundary
during the hydrogen-core burning main sequence. The entire star is
enriched in $^{4}\mathrm{He}$ (and $^{14}\mathrm{N}$ at the expense of
$^{12}\mathrm{C}$ and $^{16}\mathrm{O}$ because of the CNO cycle,
\citealt{bethe:39}), reducing the envelope opacity and preventing the
red-ward evolution and radial expansion. Nevertheless, the increase in
mean molecular weight over time leads to an increase in luminosity
\citep[e.g.,][]{kippenhahn:13, xin:2022}.

At the end of the hydrogen-core burning main sequence, each track
shows a characteristic ``hook'': this corresponds to a thermal
timescale phase of contraction because of the exhaustion of hydrogen
in the entire core region, corresponding to most of the star in these
CHE models.

Helium ignition in the center causes the star to briefly re-expand up
to its maximum radius. Nevertheless, each star remains relatively
small, with $\max(R)\lesssim 31\,R_{\odot}$ and
$T_{\rm eff}\gtrsim 10^{4.5}\,\mathrm{K}$ across the entire grid. This
minimizes the chances for binary mass-transfer with putative
companions for CHE stars \citep[e.g.,][]{demink:09,demink:16,mandel:16,marchant:16}. We emphasize that
from this point onwards, the evolution is not \emph{homogeneous}
anymore: while rotational mixing remains active, the evolutionary
timescale speeds up significantly owing to the lower energy release
per nucleon for heavier nuclear fuels, and chemical gradients seeded
by the burning impede further mixing \citep[e.g.,][]{heger:00,
  yoon:06, chieffi:13}. The stars develop a distinct carbon-oxygen
core, and an ``onion''-like structure, but lack a Hydrogen-rich,
expanded envelope.

The remaining evolution occurs roughly right-to-left at constant
luminosity, with wind mass-loss progressively revealing hotter inner
layers of the star. The hottest point and smallest radius is reached
roughly at core Neon ignition \citep{gottlieb:24}, after which stars
typically expand slightly again because of the enhanced core
contraction as neutrino cooling and nuclear neutrino losses increase.
We note that the majority of the unsuccessful models (red crosses in
\Figref{fig:grid_success}, excluded from \Figref{fig:multipanel} and
beyond) fail during or after Neon core burning. This is because this
burning phase occurs through photodisintegration
$^{20}\mathrm{Ne}(\gamma, \alpha)^{16}\mathrm{O}$ which produces
$\alpha$ particles. These have extremely high reaction rates at the
relevant temperatures $T \gtrsim 1.5\times10^{9}\,\mathrm{K}$,
\citep{kippenhahn:13}, thus the equations for the evolution of the
core composition become significantly stiffer and numerically
challenging.

The comparison between the two $40\,M_{\odot}$ and
$\omega_{\rm ZAMS}=0.6\,\omega_{\rm crit}$ models computed with a
128-isotope (thick orange solid) and 22-isotope (thin black
dot-dashed) nuclear reaction network shows how the surface $L$ and
$T_{\rm eff}$ are relatively insensitive to the treatment of very late
burning. Minor differences occur only during thermal-timescale phases
(e.g., coldest point during the ``hook'') and at the very end, when
the outer layers of these envelope-less stars can be dynamically
and/or numerically coupled to the core.

\Figref{fig:multipanel} shows an overview of their evolution and
structure at the onset of core collapse. The top left panel shows the
specific entropy profiles $s$ of the innermost $m\le4\,M_{\odot}$.
Nearly flat ``steps'' correspond to stellar regions that have
experienced \emph{efficient} convection less than a local thermal
time-scale before the onset of collapse. This results in a very nearly
adiabatic (i.e., constant $s$) profile without enough time left to
relax. The mass location of these depends on the detailed evolution of
each stellar model in our grid. The vertical axis is cut at
$s=15\,k_{B}N_{A}$ to focus on visualizing the inner core where
stellar explosions are decided, and thus excludes the outer
(higher-entropy) layers of the stars. The comparison between the solid
orange line and the thin dash-dotted black line for the two
$40\,M_{\odot}$, $\omega_{\rm ZAMS}=0.6\,\omega_{\rm crit}$ models
differing only in size of the nuclear network shows very significant
differences in the layers where explosions are decided. In this (and
other) panels, the differences introduced by the simplification of the
weak interactions is comparable or larger than the difference between
pairs of models of different mass and initial rotation computed with
large nuclear reaction networks. Thus, the systematic error introduced
by a small nuclear reaction network biases inferences about the
connection between progenitors, explosions, and remnants
\citep{farmer:16, renzo:24}.

The top right panel of \Figref{fig:multipanel} shows the profiles of
the electron fraction $Y_{e}=\sum_{j}X_{j}Z_{j}/A_{j}$, where $X_{j}$
is the mass fraction of the $j$-th isotope with charge $+Z_{j}e$ and
atomic mass $A_{j}$, in the innermost mass coordinate $m\le4\,M_{\odot}$ at the onset
of core collapse. Outside this mass coordinate $Y_e\simeq0.5$, while
inside, weak reactions deleptonize the core and lower the $Y_{e}$,
creating a profile consistent with the thermodynamics in the evolving
cores.

The bottom left panel of \Figref{fig:multipanel} depicts the full
density profiles at the onset of collapse as a function of
$\log_{10}(m/M_{\odot})$, to focus on the inner layers of our CHE
models. The inset panel shows the innermost mass coordinate
$m<5\,M_{\odot}$ on a linear scale. The density profile $\rho(m)$ in
these inner layers depends strongly on the $Y_{e}(m)$ because of the
contribution of electron degeneracy pressure to the support of the
core, and it determines the infall velocity profile once electron
captures (and later photodisintegration) remove pressure support
\citep[e.g.,][]{janka:12}. Moreover, the density determines the
initial accretion onto the shock generated at core bounce and
  thus the ram pressure impinging on the shock and
the neutrino luminosity \citep[e.g.][]{vartanyan:21, burrows:24}.
Finally, the density profile is also related to the angular momentum
profile via the mass-continuity equation relating $\rho(m)$ and the
radial distribution of mass $r(m)$. The comparison between the two
$40\,M_{\odot}$, $\omega=0.6\,\omega_{\rm crit}$ models suggests that
the collapse dynamics can be completely different in these, creating a
large systematic uncertainty on explosion models when relying on small
nuclear reaction networks \citep[e.g.,][]{renzo:24}.

The bottom right panel of \Figref{fig:multipanel} shows the specific
angular momentum per unit mass $j=r^{2}\omega$ at the onset of
core collapse. This can be compared to the angular momentum of the
innermost stable circular orbit of a putative seed BH to decide
whether a disk forms before \citep[e.g.,][]{gottlieb:24} or during a
proto-NS phase \citep[e.g.,][]{kasen:10}, or after the formation of
the BH \cite[e.g.,][]{macfadyen:94, perna:18}. With our modeling
assumptions, disk formation before BH formation appears to be more
common in our grid, meaning our models retain a large amount of
angular momentum at the onset of collapse.

\begin{figure}[tbp]
  \includegraphics[width=0.5\textwidth]{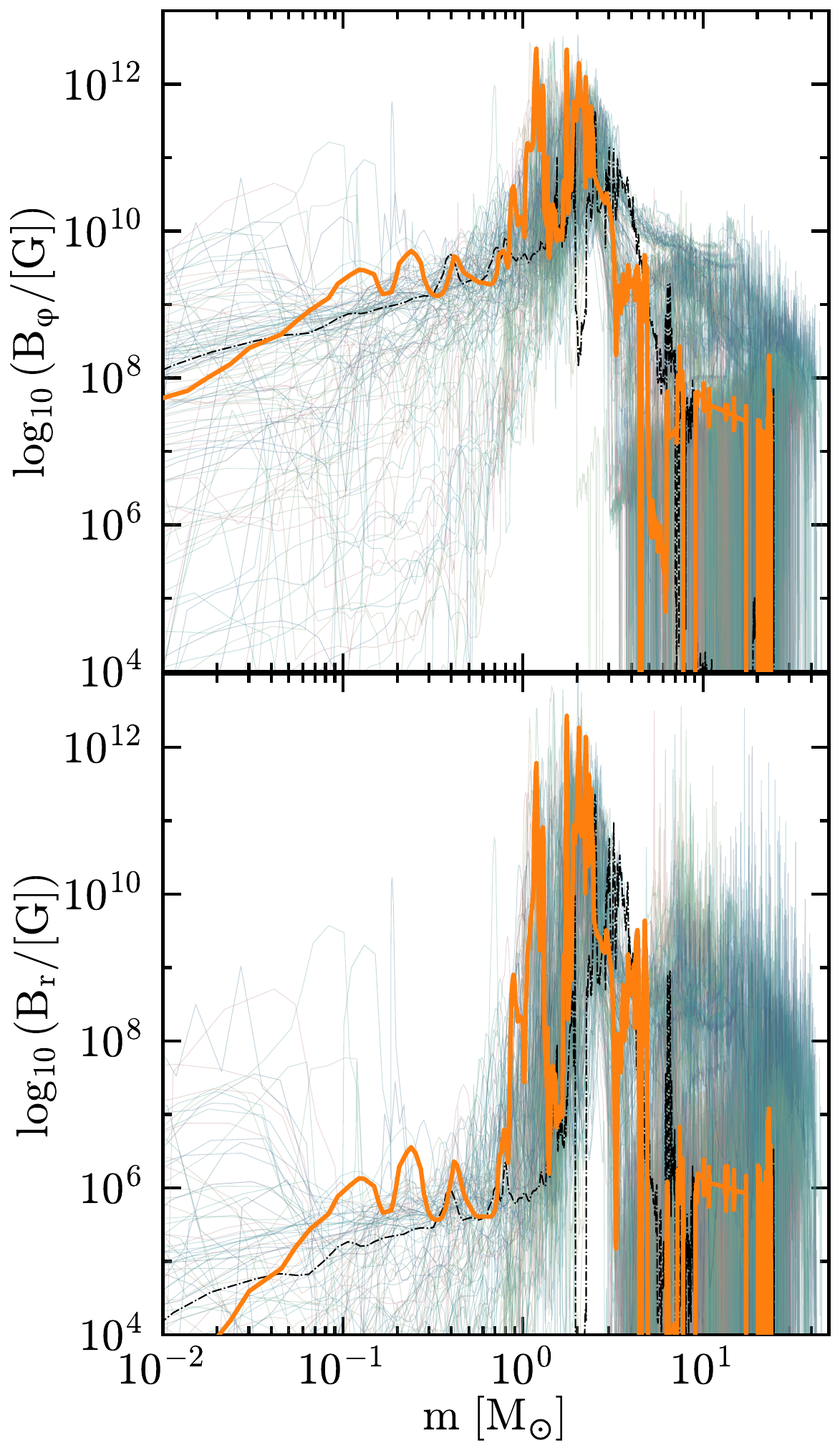}
  \caption{Azimuthal (top) and radial (bottom, local) component of the
    magnetic field generated by the Tayler-Spruit dynamo as a function
    of Lagrangian mass coordinate. The orange thick line shows
    the $40M_{\odot}$, $\omega_{\rm ZAMS}=0.6\,\omega_{\rm crit}$
    model from \cite{gottlieb:24} and the thin black dot-dashed line
    shows an otherwise identical model computed with only 22-isotopes
    from \cite{renzo:24}. Other colors indicate
      $\omega_{\rm ZAMS}/\omega_{\rm crit}$ as in
      Fig.~\ref{fig:xi_M}.}
  \label{fig:B}
\end{figure}

The Tayler-Spruit dynamo responsible for angular momentum transport in
radiative regions of our CHE models generates magnetic fields.
\Figref{fig:B} shows the azimuthal ($B_\varphi$, top) and the radial
($B_{r}$, bottom) components as a function of mass coordinate on a
logarithmic scale at the onset of collapse. The latter are
\emph{local} and not large-scale coherently organized fields. We refer
interested readers to \cite{gottlieb:24} (Sec.~2.1) for more details
and how to calculate the magnetic flux advected during the initial
phases of a core collapse.

\Figref{fig:xi_M} shows the compactness parameter \citep{oconnor:11}:
\begin{equation}
  \label{eq:compactness}
  \xi_{\mathcal{M}}=\frac{\mathcal{M}/M_{\odot}}{r(\mathcal{M})/1000\,\mathrm{km}} \ \ ,
\end{equation}
against the initial total mass $M_{\rm ZAMS}$. We use
$\mathcal{M}=1.75\,M_{\odot}$ to calculate the compactness following
\cite{burrows:24} (see also \citealt{ugliano:12}), highlighted also by
the vertical black dashed line in the top right panel of
\Figref{fig:multipanel}. Large values of $\xi_{1.75}$ are more likely
to result in BH formation, but this does not imply that the collapse
does not result in a successful explosion. In
  fact, the accretion rate on the shock, and thus the
neutrino luminosity thought to drive the shock, correlates positively
with $\xi_{1.75}$ in multi-dimensional
  simulations \cite[e.g.,][]{kuroda:18, ott:18, burrows:25}.
% The colors in \Figref{fig:xi_M} indicate the initial rotation rate.
Overall, slower rotation (darker colors) tends to result in higher
compactness, while increasing rotational support decreases the core
density slightly and decreases $\xi_{1.75}$, but the trend is not
always respected, likely due to the high non-linearity of stellar
evolution.

We report in \Tabref{tab:explodability} the values of the mass
coordinate where the specific entropy drops below $s=4\,k_{B}N_{A}$
($M_{4}$) and the mass gradient at that location
$\mu_{4}=dm/dr(m=M_{4})$ which can be used to construct a
two-parameter explodability criterion \citep{ertl:16, ertl:20}.
Unsurprisingly for the mass regime considered here, all models lie
above the critical line for BH formation, regardless of the adopted
engine calibration. Nevertheless, some of our models may retain too
much angular momentum to directly collapse into a BH.

\section{Discussion}
\label{sec:discussion}

\subsection{What astrophysical scenario can these models represent?}
\label{sec:scenario}

Direct observational evidence (e.g., from stellar spectroscopy) for
rotationally-induced CHE is still debated \cite[e.g.,][]{hunter:08,
  martins:09, martins:13,almeida:15, mahy:20, sharpe:24}, but this
evolutionary path becomes more likely at very high stellar masses and
low metallicities \citep[e.g.,][]{yoon:06}, for which samples are
sparser \cite[e.g.,][]{evans:11, shenar:24}.

To produce fast-rotating collapsar progenitors, we construct our
stellar models with initial metallicity $Z=0.001$ and \emph{assuming}
extremely fast initial rotation rates. Such initial rotations are
prohibitively rare regardless of $Z$ in young massive star populations
in the local Universe \citep[e.g.,][]{hunter:08, ramirez-agudelo:13,
  ramirez-agudelo:15, dufton:19, galan-dieguez:26, lennon:26}. While
there is growing evidence for relatively fast
($\gtrsim 250\,\mathrm{km\ s^{1}}$) primordial rotation
\citep[e.g.,][]{naze:23, britavskiy:24, naze:25}, these still have
$\omega\lesssim 0.3-0.4\,\omega_{\rm crit}$ and are not as fast as we
assume here. The fastest-rotating known O-type stars reach
$v\sin(i)\sim600\,\mathrm{km\ s^{-1}}$, likely in the regime
considered here, and are most likely accretor stars in interacting
binaries \citep{cantiello:07, ramirez-agudelo:15, renzo:21}. The
accretion process can change the core-envelope boundary in ways that
are not accounted for in our single-star models
\citep[e.g.,][]{cantiello:07, renzo:21, renzo:23, wagg:24, landri:25,
  nathaniel:25, shah:26, henneco:26}.

Nevertheless, observations of jetted explosions and LGRBs suggest that
at least some stars reach core collapse with sufficient angular
momentum to form accretion disks onto a seed compact object. This
could be the product of accretion on the secondary in an interacting
binary system \citep[e.g.,][]{cantiello:07}, stellar mergers retaining
angular momentum \citep[e.g.,][]{chatzopoulos:20,renzo:20c}, tidal
spin up in post-common envelope massive binaries
\citep{bavera:20,bavera:22,sen:25}, or possibly inefficient angular
momentum transport in stars, leading to spin up of the core as it
evolves and contracts. We do not consider here the possibility of disk
formation from stochastic angular momentum in extended convective
envelopes \citep[e.g.,][]{quataert:19, shishkin:21,antoni:25} because
our progenitors have small radii. LGRBs are associated with type Ic
broad-line SNe that require hydrogen-free progenitors with compact
envelopes. We note however that all our progenitors retain helium at
their surfaces (cf.~\Tabref{tab:explodability}), a result certainly
influenced by the assumed wind mass loss \citep[e.g.,][]{renzo:17}.

We do not attempt to investigate the detailed evolutionary scenarios
leading to fast-rotating cores. Instead, we provide a computationally
homogeneous grid of single, initially fast-rotating stellar
progenitors that could \emph{approximate} these scenarios. The grid is designed to enable systematic numerical experiments on stellar collapse
in the regimes where rotation and magnetic field are expected to play an important role.

\subsection{Known physical and numerical caveats}
\label{sec:discussion_inputs}
The rotationally-induced CHE path \citep[e.g.,][]{maeder:00,
    yoon:06} and the resulting profiles at the onset of collapse are
a product of the assumed initial rotation and the interplay between
wind-mass loss and angular momentum transport within the
  limited framework of 1D treatment of these processes
  \cite[e.g.,][]{mombarg:24,rieutord:25}. Both stellar winds and
angular momentum transport processes remain highly uncertain in
stellar context (see, e.g., \citealt{puls:08, smith:14, vink:15,
  renzo:17, josiek:24, romagnolo:26} for wind-related uncertainties,
\citealt{spruit:99, cantiello:14, fuller:19, denhartogh:20,
  skoutnev:24, skoutnev:25} for angular momentum transport
uncertainties). Other stellar physics ingredients influencing
  the size of the convective main-sequence core (e.g., overshooting),
  can also have an secondary role by changing the amount of mass that
  needs to be mixed to reach CHE.

In our spherically-symmetric models, we treat rotation in
  the ``shellular approximation'' \citep[e.g.,][]{maeder:00}, which may
  not accurately represent the multi-dimensional flows at rotation
  rates $\omega\gtrsim0.8\,\omega_{\rm crit}$. We adopt a
Tayler-Spruit dynamo in radiative regions \citep{spruit:02} for
angular momentum transport. As noted in \cite{gottlieb:24} based on
the $40\,M_{\odot}$ model with
$\omega_{\rm ZAMS}=0.6\,\omega_{\rm crit}$ (orange line in
\Figref{fig:multipanel}-\ref{fig:B}), the stronger magnetic field
saturation of the Tayler-Spruit instability suggested in
\cite{fuller:19, fuller:22} still results in CHE, but at the onset of
core collapse, the cores are too slowly rotating to produce accretion
disks. Lowering the initial rotation rate (see
  \Secref{sec:scenario}) and/or the efficiency of mixing (of
  composition and angular momentum) and/or increasing the early
  angular momentum losses could prevent rotationally-induced CHE and
  lead to a redward expansion and a clear core-envelope
  structure.

For line-driven wind mass loss, we assumed a combination of
\cite{vink:00, vink:01} and \cite{nugis:00}, as opposed to most
previous grids of fast-rotating progenitors adopting a combination of
\cite{nieuwenhuijzen:90} \citep[e.g.,][]{woosley:06} and
\cite{hamann:95} \citep[e.g.,][]{aguilera-dena:18, aguilera-dena:20}.
The lack of extended low-density hydrogen-rich envelopes
  removes the need for a ``cool'' wind mass loss, and also allows us
  to avoid known numerically challenging issues
  \citep[e.g.,][]{paxton:13, cheng:24}.

The 128-isotope nuclear reaction network we adopt (see also
Appendix~\ref{sec:mesa128}) is sufficient to resolve the density
structure of the core \citep[e.g.,][]{farmer:16}. Increasing the
network size further has a negligible impact on the electron fraction
$Y_{e}$. This has been shown both with full stellar model calculations
\citep[e.g.,][]{farmer:16} and focused one-zone experiments
\citep[e.g.,][]{grichener:25}. However, the list of isotopes adopted
here may still under-resolve the pre-collapse high-energy neutrino
flux \citep[e.g.,][]{patton:17, farag:20, myers:26} and may be
  missing important seed nuclei for the subsequent explosive
nucleosynthesis \citep[e.g.,][]{nomoto:86, hix:96, thielemann:96,
  woosley:02, yoshida:05, woosley:07, farmer:23, limongi:25}.

Moreover, the nuclear reaction rates adopted (specifically the
$^{12}\mathrm{C}(\alpha,\gamma)^{16}\mathrm{O}$ from
\citealt{kunz:02}) are under active scrutiny
\citep[e.g.,][]{deboer:17, takahashi:18, farmer:19, costa:21, woosley:21, shen:23,
  noll:25, tong:26}. Our choice of rates therefore inevitably
influences the pre-collapse structures presented here, as is true for
any stellar models in the literature \citep[see also][]{fields:18}.
Finally, we also employ the ``split-burn'' mode of \textsc{MESA}, that
is we solve separately for the composition (at fixed structure), and
iterate on the structure (at fixed updated composition). While this
approach is common in stellar evolution because of the large
difference in the timescale for nuclear burning and other processes,
it may introduce small numerical artifacts
\citep[e.g.,][]{takahashi:13, farmer:16,
  jermyn:23, limongi:24, limongi:25}.

As with any set of stellar evolution calculations, the representation
of the inherently three-dimensional convective flow
\cite[e.g.,][]{jermyn:22,joyce:23} and convective boundaries
\cite[e.g.,][]{davis:19, myers:26} are potential causes of concern.
The key hypothesis of subsonic flows holds throughout the evolution of
our models. However, our choice of free parameters ($\alpha_{\rm MLT}$
and $\alpha_{\rm ov}$) to match existing literature
\citep[e.g.,][]{brott:11, aguilera-dena:18, aguilera-dena:20} may not
be appropriate in every stellar environment
\citep[e.g.][]{joyce:18,chun:18} or for every convective layer in a
stars \citep[e.g.,][]{goldberg:22,ma:25,stuck:25,griffiths:26}, and we
do not include the latest advances from stellar hydrodynamic
simulations \citep[e.g.][]{yadav:20, schultz:20,anders:22, rizzuti:23,
  johnston:24}. Perhaps more importantly, our models adopt an
artificial strategy to expunge envelope velocities in the outer layers
which are possibly seeded by numerical noise in the fast-paced late
burning evolution. Significant efforts to improve \textsc{MESA} are
aimed at addressing this issue.
% Despite these inevitable caveats, which are inherent to numerical simulations
% that depend on input physics and physical processes under active
% investigation, our models provide a significant advance in the context
% of fast-rotating CHE stars evolved to the onset of core collapse.

\section{Conclusion}\label{sec:conclusion}

We presented a grid of rotationally-driven chemically-homogeneously
evolving massive stars at the onset of core collapse. We have
described their evolution on the Herzsprung-Russel diagram and
characterized their internal profiles at collapse,
and discussed physical and numerical limitations of our calculations.

The evolution of these initially fast-rotating models represents a
presumably extremely rare evolutionary path based on observations of
initial rotation rates in the local Universe and rates of
astrophysical transients. Our models could approximate, to a certain
degree, more common evolutionary paths where rotation is produced by
binary interactions. Since binary interactions have been shown to
introduce significant differences in progenitor structures compared to
single stellar evolution \citep[e.g.,][]{laplace:21, vartanyan:21,
  farmer:23, laplace:25, maltsev:25, ma:25b}, further studies to
compare the outcome of these pathways with single-fast rotating models
are necessary.

Despite the inevitable algorithmic limitations inherent in numerical
simulations that depend on input physics and processes under active
investigation (especially when represented by very stiff sets of
equations), our progenitor grid constitutes a significant advance in
both input physics and numerical accuracy for rapidly-rotating stars
are the onset of core collapse. By providing a homogeneous set of
pre-collapse models with substantial angular momentum, this grid
offers useful initial conditions for numerical studies of rotating
hydrogen-free massive-star explosions, including
magnetohydrodynamic-driven and jetted explosions such as collapsars
and long $\gamma$-ray bursts. We hope that our publicly available
results (at
\href{https://doi.org/10.5281/zenodo.14286306}{doi.org/10.5281/zenodo.14286306})
will be adopted in future numerical studies of the explosions
of rotating massive stars.\\

\software{This research has made use of the Astrophysics Data System,
  funded by NASA under Cooperative Agreement 80NSSC21M00561. We used
  the following software packages: \textsc{MESA} \citep{paxton:11,
    paxton:13, paxton:15, paxton:19, jermyn:23}, \texttt{python}
  \citep{python}, \texttt{numpy} \citep{numpy}, \texttt{scipy}
  \citep{2020SciPy-NMeth,scipy_8079889}, and \texttt{matplotlib}
  \citep{Hunter:2007}. Software citation information aggregated using
  \texttt{\href{https://www.tomwagg.com/software-citation-station/}{The
      Software Citation Station}}
  \citep{software-citation-station-paper,software-citation-station-zenodo}.}

\begin{acknowledgements}
  MR is grateful to N.~Britavskyi for many inspiring conversations on
  stellar rotation, and was partially supported by the US National
  Science Foundation (NSF-AST-2510584) and NASA (ATP: 95380NSSC24K0932).
  J.A.G. acknowledges financial support from NASA grant 23-ATP23-0070.
  We thank the referee for helpful feedback.
\end{acknowledgements}

\bibliographystyle{aasjournalv7}
\bibliography{CHE_GRB_progenitors.bib}

\appendix
\twocolumngrid
\section{\texttt{mesa\_128.net}}
\label{sec:mesa128}

\begin{figure}[htbp]
  \includegraphics[width=0.5\textwidth]{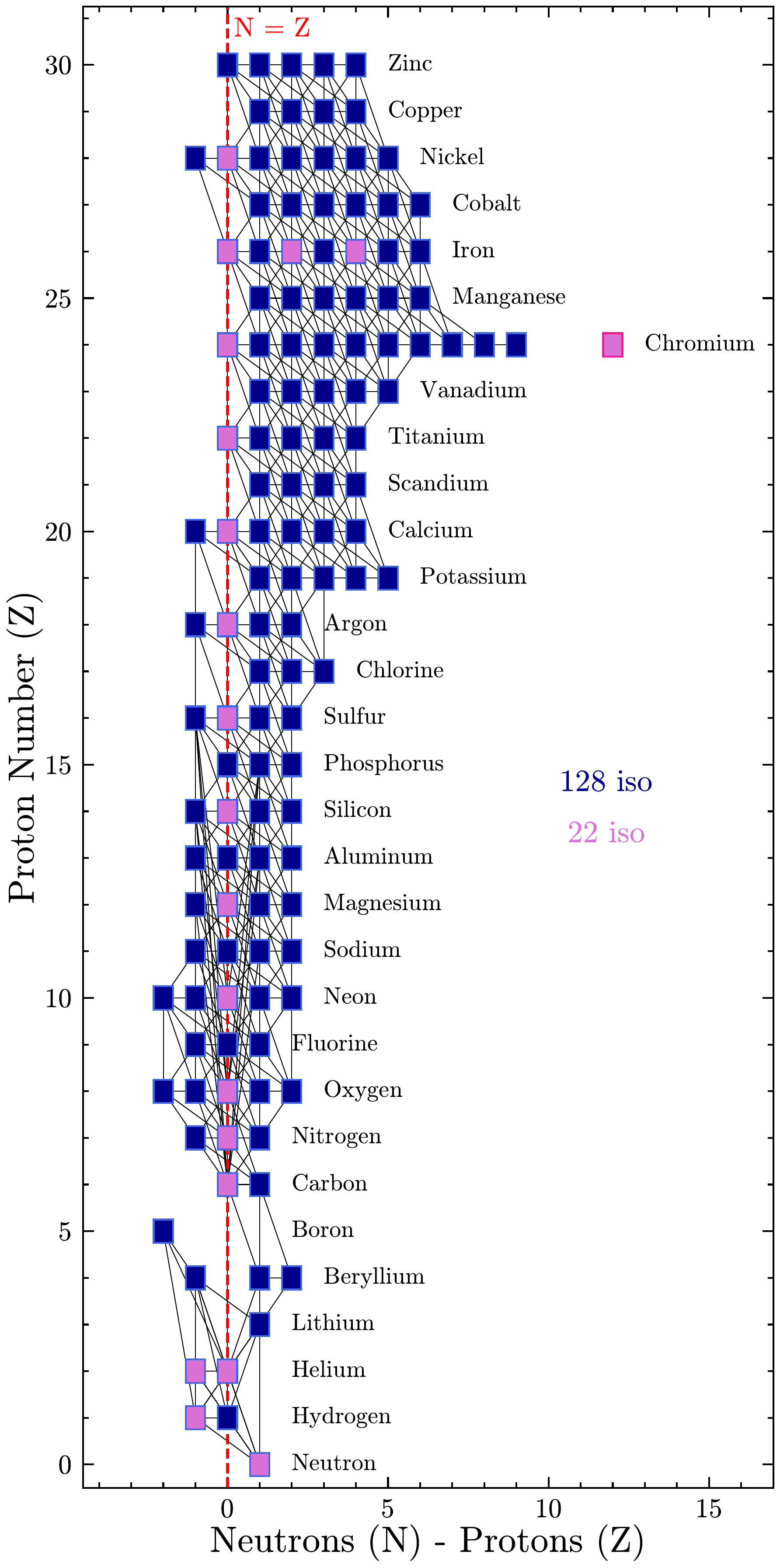}
  \caption{Isotope list and reaction links for the 128-isotope network
    used in this study and an example of 22-isotope small network.}
  \label{fig:iso}
\end{figure}

Fig.~\ref{fig:iso} shows these isotopes (in blue) with the connecting
reaction links in \texttt{mesa\_128.net} and for comparison, the
22-isotope list and reaction links of
\texttt{approx21\_cr60\_plus\_co56.net} used for our small nuclear
network model (in pink), which includes also $^{60}\mathrm{Cr}$ (not
included in \texttt{mesa\_128.net}, red outline), as an example of
isotope that can be included to tweak compound weak reactions.

The list of isotopes included in \texttt{mesa\_128.net} is:
$n$,
$^{1-2}\mathrm{H}$,
$^{2}\mathrm{H}$,
$^{3}\mathrm{He}$,
$^{4}\mathrm{He}$,
$^{7}\mathrm{Li}$,
$^{7}\mathrm{Be}$,
$^{9}\mathrm{Be}$,
$^{10}\mathrm{Be}$,
$^{8}\mathrm{B}$,
$^{12}\mathrm{C}$,
$^{13}\mathrm{C}$,
$^{13}\mathrm{N}$,
$^{14}\mathrm{N}$,
$^{15}\mathrm{N}$,
$^{14}\mathrm{O}$,
$^{15}\mathrm{O}$,
$^{16}\mathrm{O}$,
$^{17}\mathrm{O}$,
$^{18}\mathrm{O}$,
$^{17}\mathrm{F}$,
$^{18}\mathrm{F}$,
$^{19}\mathrm{F}$,
$^{18}\mathrm{Ne}$,
$^{19}\mathrm{Ne}$,
$^{20}\mathrm{Ne}$,
$^{21}\mathrm{Ne}$,
$^{22}\mathrm{Ne}$,
$^{21}\mathrm{Na}$,
$^{22}\mathrm{Na}$,
$^{23}\mathrm{Na}$,
$^{24}\mathrm{Na}$,
$^{23}\mathrm{Mg}$,
$^{24}\mathrm{Mg}$,
$^{25}\mathrm{Mg}$,
$^{26}\mathrm{Mg}$,
$^{25}\mathrm{Al}$,
$^{26}\mathrm{Al}$,
$^{27}\mathrm{Al}$,
$^{28}\mathrm{Al}$,
$^{27}\mathrm{Si}$,
$^{28}\mathrm{Si}$,
$^{29}\mathrm{Si}$,
$^{30}\mathrm{Si}$,
$^{30}\mathrm{P}$,
$^{31}\mathrm{P}$,
$^{32}\mathrm{P}$,
$^{31}\mathrm{S}$,
$^{32}\mathrm{S}$,
$^{33}\mathrm{S}$,
$^{34}\mathrm{S}$,
$^{35}\mathrm{Cl}$,
$^{36}\mathrm{Cl}$,
$^{37}\mathrm{Cl}$,
$^{35}\mathrm{Ar}$,
$^{36}\mathrm{Ar}$,
$^{37}\mathrm{Ar}$,
$^{38}\mathrm{Ar}$,
$^{39}\mathrm{K}$,
$^{40}\mathrm{K}$,
$^{41}\mathrm{K}$,
$^{42}\mathrm{K}$,
$^{43}\mathrm{K}$,
$^{39}\mathrm{Ca}$,
$^{40}\mathrm{Ca}$,
$^{41}\mathrm{Ca}$,
$^{42}\mathrm{Ca}$,
$^{43}\mathrm{Ca}$,
$^{44}\mathrm{Ca}$,
$^{43}\mathrm{Sc}$,
$^{44}\mathrm{Sc}$,
$^{45}\mathrm{Sc}$,
$^{46}\mathrm{Sc}$,
$^{44}\mathrm{Ti}$,
$^{45}\mathrm{Ti}$,
$^{46}\mathrm{Ti}$,
$^{47}\mathrm{Ti}$,
$^{48}\mathrm{Ti}$,
$^{47}\mathrm{V}$,
$^{48}\mathrm{V}$,
$^{49}\mathrm{V}$,
$^{50}\mathrm{V}$,
$^{51}\mathrm{V}$,
$^{48}\mathrm{Cr}$,
$^{49}\mathrm{Cr}$,
$^{50}\mathrm{Cr}$,
$^{51}\mathrm{Cr}$,
$^{52}\mathrm{Cr}$,
$^{53}\mathrm{Cr}$,
$^{54}\mathrm{Cr}$,
$^{55}\mathrm{Cr}$,
$^{56}\mathrm{Cr}$,
$^{57}\mathrm{Cr}$,
$^{51}\mathrm{Mn}$,
$^{52}\mathrm{Mn}$,
$^{53}\mathrm{Mn}$,
$^{54}\mathrm{Mn}$,
$^{55}\mathrm{Mn}$,
$^{56}\mathrm{Mn}$,
$^{52}\mathrm{Fe}$,
$^{53}\mathrm{Fe}$,
$^{54}\mathrm{Fe}$,
$^{55}\mathrm{Fe}$,
$^{56}\mathrm{Fe}$,
$^{57}\mathrm{Fe}$,
$^{58}\mathrm{Fe}$,
$^{55}\mathrm{Co}$,
$^{56}\mathrm{Co}$,
$^{57}\mathrm{Co}$,
$^{58}\mathrm{Co}$,
$^{59}\mathrm{Co}$,
$^{60}\mathrm{Co}$,
$^{55}\mathrm{Ni}$,
$^{56}\mathrm{Ni}$,
$^{57}\mathrm{Ni}$,
$^{58}\mathrm{Ni}$,
$^{59}\mathrm{Ni}$,
$^{60}\mathrm{Ni}$,
$^{61}\mathrm{Ni}$,
$^{59}\mathrm{Cu}$,
$^{60}\mathrm{Cu}$,
$^{61}\mathrm{Cu}$,
$^{62}\mathrm{Cu}$,
$^{60}\mathrm{Zn}$,
$^{61}\mathrm{Zn}$,
$^{62}\mathrm{Zn}$,
$^{63}\mathrm{Zn}$,
$^{64}\mathrm{Zn}$.

%%% Local Variables:
%%% mode: LaTeX
%%% TeX-master: "CHE_GRB_progenitors"
%%% End:

\section{Explodability metrics and surface composition}
\label{sec:table}

\Tabref{tab:explodability} lists for each successful model in our grid
a few key quantities related to their ``explodability'' and their
surface composition. Specifically, we list $\xi_{1.75}$ and
$M_{\rm Fe}$ shown also in \Figref{fig:xi_M}, $M_{4}$ is the outermost
mass coordinate where the specific entropy is $s=4\,k_{B}N_A$ and the
mass-gradient at that location $\mu_{4}$, which together can be
combined in the two-parameter explodability criterion by
\cite{ertl:16, ertl:20}.

The last three columns provide the surface mass fractions of the most
abundant elements, which are Helium
$Y_{\rm surf}\equiv X_{\rm surf}(^{4}\mathrm{He})$, Carbon, and
Oxygen. These do not necessarily sum to 1 because of other isotopes
present in our calculations, and are presumably sensitive to the wind
mass-loss rates and mixing processes. Interestingly, most of our
models still retain a significant amount of helium at the surface (plus,
some helium-rich material can appear in the core very late due to
photodisintegration of iron-group nuclei).
\newpage
\section{Description of the data products}
\label{sec:public_datafiles}

We share the stellar models described in this paper at
\href{https://doi.org/10.5281/zenodo.14286306}{doi.org/10.5281/zenodo.14286306},
to provide initial conditions for stellar explosion studies. In the
repository, \texttt{template.tar.xz} contains the input files for our
\textsc{MESA} simulations, designed for version \texttt{r24.03.1}.
This includes a customized \texttt{rn} script that tries to use the
\texttt{mail} command-line utility to send the last \textsc{MESA}
\texttt{pgstar} output and the last lines of the terminal output to
the user. This requires modifying the email address, and may not work
depending on the setup of \texttt{mail} on the machine running the
models. Failure to send the email notification will not affect the
\textsc{MESA} calculations in any way.

The numerical results are in \texttt{grid.tar}, which contains a
separate tarball for each star, labeled as \texttt{<M>\_rot<O>.tar.xz}
where \texttt{<M>} is the ZAMS mass in $M_{\odot}$ with one decimal
digit, and \texttt{<O>} is the initial value of
$\omega_{\rm ZAMS}/\omega_{\rm crit}$ with two decimal digits. We only
include the models reaching the onset of core collapse successfully
($v_{\rm infall}\lesssim300\,\mathrm{km\ s^{-1}}$) and pass our visual
inspection (cf.~\Figref{fig:grid_success}).

Each \texttt{<M>\_rot<O>.tar.xz} provides a \textsc{MESA} work
directory containing the input files for that specific model, the full
terminal output of the run (\texttt{output.txt}), a few plots
generated on-the-fly by \textsc{MESA} \texttt{pgstar} inside the
subfolder \texttt{png}, the \texttt{history.data} file and selected
\texttt{profile*.data} in the subfolder \texttt{LOGS}. All profiles at
the onset of core collapse are named
\texttt{CHE\_single\_core\_collapse.data}, we also provide profiles
and \textsc{MESA} \texttt{*.mod} files for the first time the
temperature in the star exceeds $\log_{10}(T/\mathrm{[K]})=8.95$
(\texttt{CHE\_logT895.mod}), core Oxygen depletion
(\texttt{O\_depl.mod}, defined as the first timestep when the central
mass fractions $X_{c}(^{16}\mathrm{O})<0.1$ and
$X_{c}(^{12}\mathrm{C})$ and $X_{c}(^{4}\mathrm{He})$ are both
$<0.001$ and $X_{c}(^{1}\mathrm{H})<0.5$) and core Silicon depletion
(\texttt{Si\_depl.mod}, defined as the first timestep when the central
mass fractions
$X_{c}(^{28}\mathrm{Si}), X_{c}(^{16}\mathrm{O}), X_{c}(^{12}\mathrm{C})<5 \times 10^{-3}$
and $X_{c}(^{4}\mathrm{He})<0.2$ and $X_{c}(^{1}\mathrm{H})<0.5$).
These conditions conservatively and consistently define the moment of
core Oxygen and Silicon core depletion across the grid with
128 isotopes.

Profiles and/or \texttt{*.mod} files can be used to initialize
multi-dimensional simulations at times earlier than the onset of core
collapse, for example, to study pre-collapse shell mergers and/or
initializing self-consistent (non-MLT) convection velocity profiles
\cite[e.g.,][]{cote:20, yadav:20, fields:22, issa:25, griffiths:26,
  issa:26}

The scripts used for the setup and analysis of our \textsc{MESA}
models are in \texttt{scripts.tar.xz}, which also contains
specifications for the dependencies. All figures can be reproduced
starting with the computed \textsc{MESA} results in \texttt{grid.tar}
(and the zenodo dataset \citep{10.5281/zenodo.11375522} from \citealt{renzo:24}
for the 22-isotope model available at \href{https://doi.org/10.5281/zenodo.11375522}{doi.org/10.5281/zenodo.11375522}).

Finally, the (interactive) \texttt{build.sh} script allows for full
reproduction of this manuscript, including downloading of the data
from their \texttt{doi}. An evolving version of the code is available
also at
\href{https://github.com/mathren/CHE_LGRB_progenitors}{github.com/mathren/CHE\_lGRB\_progenitors}.

\begin{longtable}{lc|cccc|cc|cc|ccc}
  \caption{\label{tab:explodability}\added{Core masses}, explodability metrics, and final surface composition  for our CHE models. The first two columns give the initial ZAMS values of mass and initial rotational frequency in units of $\omega_{\rm crit}$. The following three columns give the final iron ($M_{\rm Fe}$), silicon ($M_{\rm Si}$), and carbon-oxygen core masses ($M_{\rm CO}$), defined as the outermost location where the mass fraction of an element (or their sum in the case of carbon and oxygen) rises above 0.1 and the mass fraction of the lighter element drops below 0.1. For the iron core, we consider $\sum_{j} X_{j}>0.1$  where the sum
  runs over all species with $A_{j}>46$. The next two columns
  gives the Helium core mass
  $M_{\rm He}$ and total mass $M$ at terminal age main sequence, defined as $X_{c}(^{1}\mathrm{H})\le10^{-2}$. Then we list the compactness parameter $\xi_{1.75}$ \citep{oconnor:11}
  calculated at mass coordinate $\mathcal{M}=1.75\,M_{\odot}$
  \citep{burrows:24},
  and the two parameters  $M_{4}\equiv\max(m(s<4\,k_{B}N_{A}))$ and  $\mu_{4}=dm/dr(m=M_{4})$
  for the \cite{ertl:16, ertl:20} explodability criterion. The last three columns are the final surface mass
  fractions of helium
  $Y_{\rm surf}\equiv X_{\rm surf}(^{4}\mathrm{He})$, carbon, and
  oxygen. These do not necessarily sum to 1 because of other isotopes
  present in our calculations.}\\

\hline
$M_{\rm ZAMS}$ & $\omega_{\rm ZAMS}$ & $M_{\rm Fe}$ & $M_{\rm Si}$ & $M_{\rm CO}$ & $M_{\rm He, TAMS}$ & $M_{\rm TAMS}$ & \multirow{2}{*}{$\xi_{1.75}$} & $M_{4}$ & $\mu_{4}$ & \multirow{2}{*}{$Y_{\rm surf}$} & \multirow{2}{*}{$X_{\rm surf}(^{12}\mathrm{C})$} & \multirow{2}{*}{$X_{\rm surf}(^{16}\mathrm{O})$} \\
$[M_{\odot}]$ & $[\omega_{\rm crit}]$ &  $[M_{\odot}]$ & $[M_{\odot}]$ & $[M_{\odot}]$ & $[M_{\odot}]$ & $[M_{\odot}]$ &  & $[M_{\odot}]$ & $[M_{\odot}/10^{3}\,\mathrm{km}]$ & & & \\
  \hline
  \hline
  \endfirsthead
  \hline
  \multicolumn{13}{c}{(continued from previous page)}\\
  \hline
  \hline
$M_{\rm ZAMS}$ & $\omega_{\rm ZAMS}$ & $M_{\rm Fe}$ & $M_{\rm Si}$ & $M_{\rm CO}$ & $M_{\rm He, TAMS}$ & $M_{\rm TAMS}$ & \multirow{2}{*}{$\xi_{1.75}$} & $M_{4}$ & $\mu_{4}$ & \multirow{2}{*}{$Y_{\rm surf}$} & \multirow{2}{*}{$X_{\rm surf}(^{12}\mathrm{C})$} & \multirow{2}{*}{$X_{\rm surf}(^{16}\mathrm{O})$} \\
$[M_{\odot}]$ & $[\omega_{\rm crit}]$ &  $[M_{\odot}]$ & $[M_{\odot}]$ & $[M_{\odot}]$ & $[M_{\odot}]$ & $[M_{\odot}]$ &  & $[M_{\odot}]$ & $[M_{\odot}/10^{3}\,\mathrm{km}]$ & & & \\
  \hline
  \hline
  \endhead
  % python output here
\hline
30.00 & 0.55 & 1.41 & 1.41 & 19.15 & 19.20 & 24.07 & 0.592 & 1.70 & 0.173 & 0.32 & 0.16 & 0.48\\
30.00 & 0.94 & 3.14 & 1.41 & 18.37 & 17.87 & 23.01 & 0.395 & 1.56 & 0.114 & 0.26 & 0.16 & 0.51\\
\hline
32.00 & 0.55 & 2.23 & 1.48 & 20.85 & 21.93 & 26.00 & 0.544 & 1.96 & 0.134 & 0.31 & 0.21 & 0.42\\
32.00 & 0.70 & 1.76 & 1.71 & 19.33 & 19.27 & 24.61 & 0.616 & 1.81 & 0.133 & 0.31 & 0.15 & 0.46\\
32.00 & 0.84 & 2.21 & 1.34 & 19.66 & 19.14 & 24.30 & 0.468 & 1.58 & 0.143 & 0.46 & 0.09 & 0.34\\
\hline
34.00 & 0.60 & 2.10 & 1.54 & 21.59 & 20.81 & 26.29 & 0.597 & 1.88 & 0.133 & 0.66 & 0.04 & 0.22\\
34.00 & 0.65 & 1.33 & 1.32 & 20.57 & 20.63 & 26.06 & 0.650 & 1.95 & 0.179 & 0.37 & 0.14 & 0.44\\
34.00 & 0.74 & 1.45 & 1.45 & 20.53 & 20.50 & 25.67 & 0.648 & 2.05 & 0.166 & 0.24 & 0.17 & 0.54\\
34.00 & 0.84 & 1.76 & 1.76 & 20.45 & 20.40 & 25.53 & 0.651 & 1.90 & 0.132 & 0.26 & 0.16 & 0.53\\
34.00 & 0.94 & 1.46 & 1.46 & 19.73 & 20.26 & 25.44 & 0.652 & 1.91 & 0.153 & 0.40 & 0.14 & 0.41\\
34.00 & 0.99 & 2.11 & 1.49 & 20.62 & 20.30 & 25.41 & 0.545 & 1.85 & 0.127 & 0.73 & 0.04 & 0.17\\
\hline
36.00 & 0.60 & 2.10 & 1.46 & 22.01 & 22.07 & 27.53 & 0.651 & 1.90 & 0.138 & 0.26 & 0.16 & 0.50\\
36.00 & 0.65 & 3.39 & 1.46 & 21.85 & 21.89 & 27.23 & 0.542 & 1.76 & 0.186 & 0.28 & 0.16 & 0.49\\
36.00 & 0.70 & 2.11 & 1.52 & 21.69 & 21.83 & 27.05 & 0.683 & 1.92 & 0.144 & 0.25 & 0.17 & 0.50\\
36.00 & 0.74 & 1.49 & 1.50 & 21.79 & 21.74 & 26.92 & 0.631 & 2.08 & 0.180 & 0.82 & 0.03 & 0.10\\
36.00 & 0.79 & 2.20 & 1.56 & 21.88 & 21.62 & 26.81 & 0.551 & 1.99 & 0.134 & 0.75 & 0.04 & 0.15\\
36.00 & 0.84 & 1.31 & 1.31 & 20.93 & 21.56 & 26.67 & 0.640 & 2.14 & 0.193 & 0.29 & 0.17 & 0.51\\
36.00 & 0.89 & 1.42 & 1.42 & 21.26 & 21.49 & 26.66 & 0.651 & 2.11 & 0.180 & 0.25 & 0.17 & 0.53\\
36.00 & 0.94 & 2.13 & 1.51 & 21.17 & 21.58 & 26.64 & 0.615 & 1.94 & 0.134 & 0.29 & 0.16 & 0.47\\
36.00 & 0.99 & 2.12 & 1.45 & 21.32 & 21.56 & 26.57 & 0.530 & 1.91 & 0.131 & 0.52 & 0.07 & 0.30\\
\hline
38.00 & 0.60 & 2.18 & 1.64 & 23.39 & 23.32 & 28.79 & 0.696 & 2.12 & 0.180 & 0.76 & 0.05 & 0.14\\
38.00 & 0.70 & 2.36 & 1.43 & 22.65 & 23.06 & 28.25 & 0.559 & 1.77 & 0.148 & 0.46 & 0.08 & 0.34\\
38.00 & 0.74 & 2.45 & 1.45 & 22.81 & 23.04 & 28.09 & 0.536 & 1.69 & 0.150 & 0.61 & 0.07 & 0.23\\
38.00 & 0.79 & 1.66 & 1.66 & 22.67 & 22.89 & 28.00 & 0.661 & 2.16 & 0.229 & 0.70 & 0.06 & 0.18\\
38.00 & 0.84 & 1.41 & 1.41 & 22.35 & 22.80 & 27.89 & 0.642 & 2.07 & 0.197 & 0.70 & 0.06 & 0.17\\
38.00 & 0.89 & 1.54 & 1.54 & 21.96 & 22.80 & 27.84 & 0.604 & 1.74 & 0.150 & 0.53 & 0.10 & 0.30\\
38.00 & 0.94 & 1.42 & 1.42 & 22.58 & 22.80 & 27.76 & 0.640 & 1.98 & 0.177 & 0.67 & 0.07 & 0.19\\
38.00 & 0.99 & 2.16 & 1.54 & 22.27 & 22.73 & 27.76 & 0.575 & 1.99 & 0.121 & 0.48 & 0.09 & 0.32\\
\hline
40.00 & 0.55 & 1.48 & 1.48 & 24.38 & 24.74 & 30.46 & 0.588 & 1.81 & 0.204 & 0.28 & 0.17 & 0.50\\
40.00 & 0.60 & 1.62 & 1.62 & 24.36 & 24.63 & 29.94 & 0.656 & 2.10 & 0.177 & 0.85 & 0.06 & 0.07\\
40.00 & 0.65 & 2.76 & 1.65 & 23.94 & 24.50 & 29.62 & 0.637 & 1.81 & 0.158 & 0.83 & 0.07 & 0.08\\
40.00 & 0.70 & 2.23 & 1.51 & 23.58 & 24.38 & 29.39 & 0.637 & 1.94 & 0.135 & 0.45 & 0.10 & 0.33\\
40.00 & 0.79 & 2.18 & 1.55 & 23.47 & 24.15 & 29.08 & 0.474 & 1.94 & 0.114 & 0.27 & 0.12 & 0.45\\
40.00 & 0.84 & 2.16 & 1.52 & 23.24 & 24.06 & 28.99 & 0.541 & 1.80 & 0.146 & 0.55 & 0.10 & 0.27\\
40.00 & 0.89 & 2.68 & 23.82 & 23.34 & 24.08 & 28.96 & 0.609 & 1.61 & 0.218 & 0.90 & 0.07 & 0.03\\
40.00 & 0.94 & 1.49 & 1.49 & 23.43 & 23.99 & 28.91 & 0.666 & 2.07 & 0.178 & 0.71 & 0.09 & 0.16\\
40.00 & 0.99 & 1.37 & 1.37 & 23.40 & 24.01 & 28.90 & 0.665 & 2.08 & 0.181 & 0.88 & 0.07 & 0.04\\
\hline
42.00 & 0.60 & 1.72 & 1.72 & 25.12 & 26.00 & 31.06 & 0.698 & 2.00 & 0.186 & 0.73 & 0.11 & 0.14\\
42.00 & 0.65 & 1.59 & 1.59 & 24.82 & 25.74 & 30.80 & 0.602 & 2.14 & 0.147 & 0.84 & 0.11 & 0.05\\
42.00 & 0.70 & 2.07 & 1.77 & 24.53 & 25.67 & 30.53 & 0.602 & 1.88 & 0.135 & 0.34 & 0.15 & 0.41\\
42.00 & 0.74 & 1.81 & 1.81 & 24.57 & 25.48 & 30.44 & 0.684 & 1.88 & 0.178 & 0.74 & 0.11 & 0.12\\
42.00 & 0.79 & 1.89 & 1.89 & 24.39 & 25.48 & 30.24 & 0.671 & 2.18 & 0.211 & 0.77 & 0.11 & 0.10\\
42.00 & 0.84 & 2.16 & 1.40 & 24.39 & 25.36 & 30.13 & 0.464 & 1.74 & 0.143 & 0.44 & 0.12 & 0.33\\
42.00 & 0.89 & 1.58 & 1.58 & 24.16 & 25.29 & 30.03 & 0.642 & 2.04 & 0.197 & 0.78 & 0.11 & 0.09\\
42.00 & 0.94 & 2.32 & 1.45 & 24.25 & 25.31 & 30.03 & 0.527 & 1.72 & 0.162 & 0.67 & 0.11 & 0.17\\
42.00 & 0.99 & 2.08 & 1.58 & 23.77 & 25.31 & 30.01 & 0.613 & 1.91 & 0.138 & 0.44 & 0.13 & 0.33\\
\hline
44.00 & 0.60 & 2.18 & 1.50 & 26.07 & 27.27 & 32.31 & 0.549 & 1.77 & 0.150 & 0.37 & 0.15 & 0.38\\
44.00 & 0.65 & 2.17 & 1.46 & 25.58 & 27.12 & 31.93 & 0.579 & 1.93 & 0.131 & 0.40 & 0.15 & 0.35\\
44.00 & 0.70 & 2.41 & 1.00 & 25.47 & 26.94 & 31.67 & 0.493 & 1.68 & 0.140 & 0.63 & 0.15 & 0.18\\
44.00 & 0.74 & 1.37 & 1.37 & 25.31 & 26.79 & 31.51 & 0.645 & 2.07 & 0.190 & 0.79 & 0.15 & 0.06\\
44.00 & 0.79 & 2.21 & 1.49 & 25.22 & 26.69 & 31.35 & 0.497 & 1.95 & 0.134 & 0.46 & 0.15 & 0.31\\
44.00 & 0.84 & 2.02 & 2.02 & 25.02 & 26.62 & 31.27 & 0.661 & 2.13 & 0.236 & 0.78 & 0.15 & 0.07\\
44.00 & 0.89 & 2.77 & 1.33 & 25.17 & 26.54 & 31.16 & 0.674 & 1.74 & 0.205 & 0.69 & 0.15 & 0.14\\
44.00 & 0.94 & 1.44 & 1.44 & 25.05 & 26.49 & 31.14 & 0.580 & 1.95 & 0.148 & 0.38 & 0.15 & 0.37\\
44.00 & 0.99 & 2.10 & 1.58 & 24.98 & 26.52 & 31.12 & 0.622 & 1.87 & 0.145 & 0.39 & 0.15 & 0.35\\
\hline
46.00 & 0.55 & 1.75 & 1.75 & 27.21 & 28.91 & 33.86 & 0.531 & 1.94 & 0.141 & 0.53 & 0.17 & 0.24\\
46.00 & 0.60 & 1.95 & 1.95 & 26.89 & 28.53 & 33.42 & 0.664 & 2.05 & 0.226 & 0.73 & 0.18 & 0.08\\
46.00 & 0.65 & 2.83 & 1.89 & 26.33 & 28.38 & 33.06 & 0.665 & 1.96 & 0.199 & 0.75 & 0.18 & 0.07\\
46.00 & 0.89 & 2.21 & 1.81 & 25.99 & 27.78 & 32.27 & 0.610 & 1.88 & 0.148 & 0.40 & 0.17 & 0.34\\
\hline
48.00 & 0.84 & 1.39 & 1.39 & 26.52 & 29.07 & 33.47 & 0.602 & 1.92 & 0.164 & 0.54 & 0.21 & 0.21\\
48.00 & 0.94 & 1.38 & 1.38 & 26.62 & 28.98 & 33.29 & 0.640 & 1.88 & 0.177 & 0.47 & 0.19 & 0.27\\
\hline
50.00 & 0.55 & 1.50 & 1.50 & 28.42 & 31.43 & 36.07 & 0.676 & 2.01 & 0.199 & 0.62 & 0.26 & 0.12\\
50.00 & 0.60 & 1.26 & 1.26 & 28.42 & 31.15 & 35.62 & 0.579 & 2.03 & 0.149 & 0.54 & 0.23 & 0.19\\
50.00 & 0.70 & 2.33 & 1.02 & 28.05 & 30.68 & 34.98 & 0.545 & 1.58 & 0.176 & 0.52 & 0.22 & 0.22\\
50.00 & 0.79 & 2.30 & 1.40 & 27.90 & 30.39 & 34.66 & 0.574 & 1.89 & 0.154 & 0.56 & 0.23 & 0.18\\
50.00 & 0.84 & 1.36 & 1.36 & 27.75 & 30.28 & 34.54 & 0.506 & 1.62 & 0.155 & 0.52 & 0.22 & 0.21\\
\hline
52.00 & 0.50 & 1.59 & 1.59 & 30.15 & 33.28 & 37.72 & 0.653 & 2.04 & 0.201 & 0.57 & 0.29 & 0.13\\
52.00 & 0.55 & 1.60 & 1.60 & 29.68 & 32.77 & 37.20 & 0.645 & 2.23 & 0.204 & 0.59 & 0.28 & 0.12\\
52.00 & 0.60 & 1.81 & 1.81 & 29.22 & 32.39 & 36.73 & 0.688 & 2.20 & 0.199 & 0.61 & 0.27 & 0.11\\
52.00 & 0.65 & 2.02 & 2.01 & 28.88 & 32.16 & 36.32 & 0.660 & 2.26 & 0.263 & 0.63 & 0.26 & 0.11\\
52.00 & 0.84 & 2.31 & 1.53 & 28.37 & 31.54 & 35.55 & 0.631 & 2.00 & 0.153 & 0.40 & 0.23 & 0.31\\
\hline
54.00 & 0.79 & 1.79 & 1.79 & 29.32 & 32.81 & 36.74 & 0.730 & 1.97 & 0.194 & 0.37 & 0.27 & 0.32\\
54.00 & 0.84 & 2.09 & 2.09 & 29.38 & 32.69 & 36.61 & 0.658 & 2.18 & 0.248 & 0.31 & 0.27 & 0.36\\
54.00 & 0.94 & 1.48 & 1.47 & 29.27 & 32.47 & 36.48 & 0.655 & 1.96 & 0.195 & 0.54 & 0.27 & 0.17\\
\hline
56.00 & 0.65 & 2.81 & 2.46 & 30.73 & 34.60 & 38.47 & 0.667 & 2.49 & 0.277 & 0.29 & 0.29 & 0.37\\
56.00 & 0.74 & 2.12 & 2.12 & 30.40 & 34.16 & 38.00 & 0.649 & 2.29 & 0.260 & 0.38 & 0.29 & 0.30\\
56.00 & 0.89 & 2.09 & 2.09 & 29.98 & 33.80 & 37.54 & 0.661 & 2.24 & 0.249 & 0.39 & 0.29 & 0.29\\
56.00 & 0.94 & 1.42 & 1.42 & 30.01 & 33.71 & 37.50 & 0.644 & 1.96 & 0.185 & 0.44 & 0.27 & 0.25\\
56.00 & 0.99 & 2.07 & 2.07 & 29.98 & 33.74 & 37.45 & 0.621 & 2.34 & 0.268 & 0.45 & 0.30 & 0.24\\
\hline
58.00 & 0.55 & 2.33 & 1.69 & 32.09 & 36.61 & 40.38 & 0.646 & 2.24 & 0.186 & 0.54 & 0.33 & 0.13\\
58.00 & 0.65 & 2.11 & 2.11 & 31.62 & 35.81 & 39.51 & 0.642 & 2.63 & 0.340 & 0.42 & 0.31 & 0.24\\
58.00 & 0.74 & 2.01 & 2.01 & 31.10 & 35.41 & 38.99 & 0.666 & 2.26 & 0.268 & 0.42 & 0.32 & 0.25\\
58.00 & 0.94 & 2.51 & 1.99 & 30.91 & 34.88 & 38.49 & 0.656 & 2.33 & 0.248 & 0.39 & 0.30 & 0.28\\
58.00 & 0.99 & 1.61 & 1.61 & 30.62 & 34.84 & 38.44 & 0.679 & 2.14 & 0.229 & 0.56 & 0.31 & 0.12\\
\hline
60.00 & 0.50 & 2.01 & 2.01 & 33.63 & 38.53 & 41.95 & 0.626 & 2.51 & 0.293 & 0.36 & 0.32 & 0.28\\
60.00 & 0.60 & 2.27 & 2.27 & 32.82 & 37.38 & 40.97 & 0.604 & 2.54 & 0.294 & 0.48 & 0.33 & 0.18\\
60.00 & 0.94 & 2.15 & 2.15 & 31.70 & 36.04 & 39.48 & 0.602 & 2.59 & 0.305 & 0.52 & 0.33 & 0.15\\
\hline
62.00 & 0.50 & 2.17 & 2.17 & 34.44 & 39.75 & 42.98 & 0.603 & 2.64 & 0.297 & 0.42 & 0.32 & 0.22\\
62.00 & 0.60 & 2.04 & 2.04 & 33.61 & 38.62 & 42.00 & 0.658 & 2.40 & 0.282 & 0.39 & 0.33 & 0.25\\
62.00 & 0.70 & 2.04 & 2.04 & 33.00 & 37.92 & 41.28 & 0.633 & 2.38 & 0.276 & 0.38 & 0.33 & 0.27\\
\hline
64.00 & 0.74 & 2.12 & 2.12 & 33.58 & 38.90 & 42.03 & 0.607 & 2.41 & 0.278 & 0.38 & 0.33 & 0.26\\
64.00 & 0.84 & 2.43 & 2.43 & 33.33 & 38.53 & 41.63 & 0.589 & 2.76 & 0.352 & 0.37 & 0.33 & 0.27\\
\hline
66.00 & 0.60 & 2.07 & 2.07 & 35.34 & 41.02 & 43.99 & 0.627 & 2.61 & 0.343 & 0.45 & 0.36 & 0.18\\
66.00 & 0.94 & 2.15 & 2.15 & 33.91 & 39.46 & 42.40 & 0.609 & 2.43 & 0.315 & 0.38 & 0.33 & 0.26\\
\hline
68.00 & 0.65 & 2.17 & 2.17 & 35.69 & 41.80 & 44.55 & 0.598 & 2.58 & 0.318 & 0.40 & 0.35 & 0.23\\
\hline
70.00 & 0.60 & 2.14 & 2.14 & 36.93 & 43.47 & 45.97 & 0.606 & 2.60 & 0.294 & 0.45 & 0.37 & 0.17\\
\hline
72.00 & 0.74 & 2.18 & 2.18 & 36.80 & 43.55 & 45.90 & 0.594 & 2.61 & 0.311 & 0.37 & 0.35 & 0.25\\
72.00 & 0.79 & 2.04 & 2.04 & 36.63 & 43.30 & 45.64 & 0.661 & 2.58 & 0.307 & 0.49 & 0.38 & 0.13\\
72.00 & 0.89 & 2.29 & 2.29 & 36.28 & 42.92 & 45.29 & 0.599 & 2.57 & 0.313 & 0.46 & 0.38 & 0.15\\
\hline
74.00 & 0.60 & 2.14 & 2.14 & 38.41 & 45.87 & 47.88 & 0.644 & 2.64 & 0.309 & 0.48 & 0.39 & 0.13\\
74.00 & 0.94 & 2.12 & 2.12 & 36.96 & 43.94 & 46.11 & 0.657 & 2.31 & 0.274 & 0.49 & 0.38 & 0.13\\
\hline
76.00 & 0.55 & 2.46 & 2.21 & 39.38 & 47.62 & 49.38 & 0.644 & 2.32 & 0.253 & 0.47 & 0.39 & 0.13\\
76.00 & 0.74 & 1.49 & 1.49 & 38.17 & 45.89 & 47.75 & 0.645 & 1.98 & 0.235 & 0.48 & 0.39 & 0.12\\
\hline
78.00 & 0.74 & 2.02 & 2.02 & 38.93 & 47.00 & 48.68 & 0.646 & 2.58 & 0.350 & 0.44 & 0.38 & 0.17\\
78.00 & 0.79 & 1.99 & 1.99 & 38.69 & 46.73 & 48.41 & 0.645 & 2.23 & 0.238 & 0.44 & 0.37 & 0.17\\
\hline
80.00 & 0.50 & 2.10 & 2.10 & 41.50 & 51.01 & 52.04 & 0.669 & 2.32 & 0.273 & 0.48 & 0.40 & 0.12\\
\hline
82.00 & 0.50 & 2.26 & 2.26 & 42.27 & 52.30 & 52.98 & 0.603 & 2.62 & 0.361 & 0.45 & 0.39 & 0.15\\
\hline
84.00 & 0.55 & 2.37 & 2.37 & 42.31 & 52.53 & 53.15 & 0.604 & 2.87 & 0.385 & 0.47 & 0.40 & 0.12\\
84.00 & 0.60 & 2.25 & 2.25 & 42.05 & 51.83 & 52.54 & 0.593 & 2.68 & 0.369 & 0.34 & 0.35 & 0.27\\
84.00 & 0.70 & 2.40 & 2.40 & 41.23 & 50.78 & 51.65 & 0.590 & 2.75 & 0.386 & 0.36 & 0.36 & 0.25\\
84.00 & 0.74 & 2.27 & 2.27 & 41.13 & 50.49 & 51.38 & 0.605 & 2.72 & 0.378 & 0.48 & 0.40 & 0.12\\
\hline
88.00 & 0.84 & 2.64 & 1.55 & 41.88 & 52.08 & 52.63 & 0.642 & 2.38 & 0.266 & 0.48 & 0.40 & 0.12\\
\hline
96.00 & 0.70 & 2.28 & 2.28 & 44.98 & 56.94 & 56.94 & 0.587 & 2.66 & 0.352 & 0.48 & 0.41 & 0.11\\
  \hline
\end{longtable}

%%% Local Variables:
%%% mode: LaTeX
%%% TeX-master: "CHE_GRB_progenitors"
%%% End:

\end{document}